\documentclass[11pt,twoside]{./atmp}

\usepackage{amsmath,amssymb,amsfonts}
\usepackage{graphicx}
\usepackage[square,comma,numbers,sort&compress]{natbib}

\usepackage[all]{xy}
\usepackage{color}

\def\R{{\mathbb R}}
\def\y{{\bf y}}

\def\x{{\bf x}}
\def\e{{\bf e}}
\def\n{{\bf n}}

\def\dc{\partial}
\def\pe{\perp}
\def\oe{\overline}

\begin{document}

\title[Exponential decay of Laplacian eigenfunctions]
{Exponential decay of Laplacian eigenfunctions in domains with branches}


\author[A. Delitsyn, B.-T. Nguyen, D. S. Grebenkov]{Andrey Delitsyn$^1$, Binh-Thanh Nguyen$^2$, \\ and Denis S. Grebenkov$^{2-4}$}

\address{
$^1$ Mathematical Department of the Faculty of Physics, Moscow State University, 119991 Moscow, Russia \\
$^2$ Laboratoire de Physique de la Mati\`{e}re Condens\'{e}e (UMR 7643), \\
CNRS -- Ecole Polytechnique, F-91128 Palaiseau, France \\
$^3$ Laboratoire Poncelet, CNRS -- Independent University of Moscow, \\  
Bolshoy Vlasyevskiy Pereulok 11, 119002 Moscow, Russia \\
$^4$ Chebyshev Laboratory, Saint Petersburg State University, \\
14th line of Vasil'evskiy Ostrov 29, Saint Petersburg, Russia }

\addressemail{delitsyn@mail.ru, binh-thanh.nguyen@polytechnique.edu, denis.grebenkov@polytechnique.edu}

\begin{abstract}
The behavior of Laplacian eigenfunctions in domains with branches is
investigated.  If an eigenvalue is below a threshold which is
determined by the shape of the branch, the associated eigenfunction is
proved to exponentially decay inside the branch.  The decay rate is
twice the square root of the difference between the threshold and the
eigenvalue.  The derived exponential estimate is applicable for
arbitrary domains in any spatial dimension.  Numerical simulations
illustrate and further extend the theoretical estimate.
\end{abstract}



\maketitle

\section{ Introduction }

The Laplace operator eigenfunctions play a crucial role in different
fields of physics: vibration modes of a thin membrane, standing waves
in optical or acoustical cavity resonators, the natural spectral
decomposition basis for diffusive processes, the eigenstates of a
single trapped particle in quantum mechanics, etc.  For the unit
interval, the spectrum of the Laplace operator is particularly simple,
and the eigenfunctions are just linear combinations of Fourier
harmonics $\{e^{i\pi nx}\}$.  The oscillating character of
eigenfunctions is then often expected for domains in two and higher
dimensions.  The ``scholar'' examples of the explicit eigenbases in
rectangular and circular domains strongly support this oversimplified
but common view.  At the same time, the geometrical structure of
eigenfunctions may be extremely complicated even for simple domains
(e.g., the structure of nodal lines of degenerate eigenfunctions in a
square \cite{Courant}).  The more complex the domain is, the more
sophisticated and sometimes unexpected the behavior of the associated
eigenfunctions may be.  For instance, numerous numerical and
experimental studies of the eigenvalue problem in irregular or
(pre)fractal domains revealed the existence of weakly localized
eigenfunctions which have pronounced amplitudes only on small
subregions of the domain
\cite{Haeberle98,Sapoval97,Hebert99,Felix07,Sapoval93,Russ97,Russ02,Filoche09,Sapoval91,Even99,Heilman10,Daudert07}.

In this paper, we investigate the behavior of Laplacian eigenfunctions
in domains with branches.  We show that certain eigenfunctions are
``expelled'' from the branch, i.e., their amplitude along the branch
decays exponentially fast.  A similar ``expulsion'' effect is well
known in optics and acoustics: a wave of wavelength $\ell$ cannot
freely propagate inside a rectangular channel of width $b$ smaller
than $\ell/2$ because of the exponential attenuation $\sim
e^{-x\pi/b}$ along the channel \cite{Sapoval91,Jackson}.  We extend
this classical result to arbitrary domains with branches of arbitrary
shape.  We derive a rigorous exponential estimate for the $L_2$-norm
of the eigenfunction in cross-sections of the branch.  We obtain the
sharp decay rate which generalizes and refines the classical rate
$\pi/b$.  Although this problem is remotely related to localization of
waves in optical or acoustical waveguides (e.g., infinite bended tubes
\cite{Jackson,Goldstone92,Carini93}), we mainly focus on bounded
domains.

It is worth noting that the exponential decay of eigenfunctions of
Schr\"odinger operators in free space (so-called strong localization)
has been thoroughly investigated in physical and mathematical
literature (see, e.g.,
\cite{Schnol57,Orocko74,Agmon,Maslov64,Maslov}).  The first
exploration of this problem for arbitrary Schr\"odinger potential
bounded from below was given by Schnol' \cite{Schnol57} who proved an
exponential decay of eigenfunctions in which the decay rate was
related to the distance between the corresponding eigenvalue and the
essential spectrum.  This result is of remarkable generality because
the essential spectrum may be arbitrary, for example with gaps, and
may not consist of positive axis.  A sharp estimate for the decay rate
was made by Maslov who reduced the problem to a differential
inequality \cite{Maslov64,Maslov}.  Anderson discovered the
exponential decay of eigenfunctions of the Schr\"odinger operator with
random potentials \cite{Anderson58}.  This phenomenon, known as the
Anderson localization, has been intensively investigated (see reviews
\cite{Belitz94,Evers08}).  Although the Laplace operator in a bounded
domain is a much simpler mathematical object, the properties of its
eigenfunctions are still poorly understood.

The paper is organized as follows.  In Sec. \ref{sec:rectangular}, we
start by considering a two-dimensional domain with a rectangular
branch.  In this special case, the estimates are derived in a rather
elementary and straightforward way that helps to illustrate many
properties of eigenfunctions.  Section \ref{sec:general} presents the
analysis for domains with branches of arbitrary shape in any spatial
dimension.  We provide a sufficient condition on the eigenvalue, under
which the related eigenfunction decays exponentially inside the
branch.  Sec. \ref{sec:discussion} presents numerical examples which
illustrate the theoretical results and suggest new perspectives for
further investigations.  The paper ends by conclusions, in which the
main results are summarized and their consequences are discussed.

\section{ Rectangular branch }
\label{sec:rectangular}

\begin{figure}
\begin{center}
\includegraphics[width=100mm]{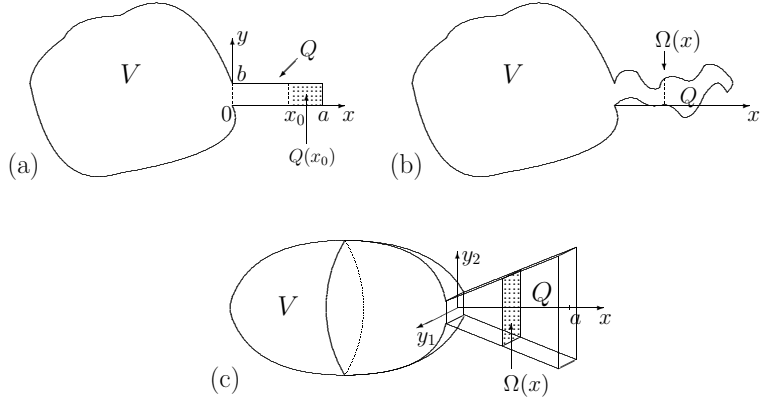}
\end{center}
\caption{
A bounded domain $D$ is the union of a basic domain $V$ of arbitrary
shape and a branch $Q$.  Our goal is to estimate how an eigenfunction
of the Laplace operator in $D$ decays inside the branch.  {\bf (a)}
rectangular branch $Q = \{ (x,y)\in \R^2~:~ 0<x<a,~0<y<b\}$ (the
energetic norm is estimated in the subdomain $Q(x_0)$ shown by
shadowed region); {\bf (b)} branch of arbitrary shape [the $L_2$-norm
is estimated in the cross-section $\Omega(x)$]; {\bf (c)}
three-dimensional increasing branch. }
\label{fig:domain}
\end{figure}

We begin with the following example.  We consider the Dirichlet
eigenvalue problem
\begin{equation}
\label{1}
 - \Delta u = \lambda u , \quad (x,y) \in D,  \quad \quad u|_{\dc D} = 0  
\end{equation}
in a planar bounded domain $ \oe D = \oe V \cup \oe Q $ which is
decomposed into a basic domain $V$ of arbitrary shape and a
rectangular branch $Q$ with sides $a$ and $b$: $Q = \{ (x,y)\in\R^2~:~
0 < x < a,~ 0 < y < b \}$, as illustrated on Fig. \ref{fig:domain}a.
We assume that the eigenvalue $\lambda$ is smaller than the first
eigenvalue of the Laplace operator in the cross-section of the branch
(i.e., the interval $[0,b]$):
\begin{equation}
\label{eq:assumpt0}
\lambda < \pi^2/b^2 .
\end{equation}
Under this condition, we aim to show the exponential decay of the
associated eigenfunction $ u $ in the branch $Q$.

In this illustrative example, the derivation is based on an explicit
upper bound for the norm of the eigenfunction $u$.  One can easily
check that a general solution of Eq. (\ref{1}) in the rectangular
branch $Q$ has a form
\begin{equation}
\label{3}
u(x,y) = \sum \limits_{n=1}^{\infty} c_n \sinh(\gamma_n (a-x)) \sin(\pi n y/b) , 
\end{equation}
where $\gamma_n = \sqrt{(\frac{\pi}{b} n)^2 - \lambda}$, and $c_n$ are
constants.  We consider the energetic norm of the function $u$ in the
subdomain $Q(x_0) = \{ (x,y)\in\R^2~:~ x_0 < x < a,~ 0 < y < b \}$
which is defined as
\begin{equation}
\label{4}
||\nabla u||^2_{L_2(Q(x_0))} \equiv \int \limits_{Q(x_0)} (\nabla u, \nabla u) dx dy . 
\end{equation}
Substituting Eq. (\ref{3}) to (\ref{4}) yields
\begin{equation} 
\label{eq:energy}
\begin{split}
||\nabla u||^2_{L_2(Q(x_0))} & = \sum\limits_{n=1}^{\infty} c_n^2 \frac{b}{2} \int\limits_{x_0}^a 
\biggl[(\frac{\pi}{b}n)^2 \sinh^2 (\gamma_n (a - x)) \\ 
& + \gamma^2_n \cosh^2(\gamma_n (a - x))\biggr] dx  . \\
\end{split}
\end{equation}
Using elementary inequalities for the integral (see Appendix
\ref{sec:rectangle}), one gets
\begin{equation}
\label{eq:aux10}
||\nabla u||^2_{L_2(Q(x_0))} \leq C e^{-2\gamma_1 x_0} \sum\limits_{n=1}^{\infty} n c_n^2 \sinh^2(\gamma_n a) ,
\end{equation}
where $C$ is an explicit constant.  The trace theorem provides the
upper bound for the above series (see Appendix
\ref{sec:rectangle})
\begin{equation*}
||\nabla u||^2_{L_2(Q(x_0))} \leq C_1 e^{-2\gamma_1 x_0} ,
\end{equation*}
with another explicit constant $C_1$.  So, we established the
exponential decay of the energy $||\nabla u||^2$ in the branch $Q$
with the decay rate $2\gamma_1 = 2\sqrt{(\frac{\pi}{b})^2 - \lambda}$.

From this estimate it is easy to deduce a similar estimate in the
$L_2$-norm,
\begin{equation}
\label{eq:decay_rect}
||u||^2_{L_2(Q(x_0))} \equiv \int\limits_{Q(x_0)} u^2 dx dy \leq C_2 e^{-2\gamma_1 x_0}  ,
\end{equation}
where $C_2$ is another constant.  The derivation implies that the
estimate is sharp, i.e. the decay rate cannot be improved in general.

It is worth noting that no information about the basic domain $V$ was
used.  In particular, the Dirichlet boundary condition on $\dc V$ can
be replaced by arbitrary boundary condition under which the Laplace
operator is still self-adjoint.

\section{ Branch of arbitrary shape }
\label{sec:general}

Now we show that the above estimate remains valid for eigenfunctions
in a much more general case with a branch of arbitrary shape in
$\R^{n+1}$ ($n = 1,2,3,...$).  We consider the eigenvalue problem
\begin{equation}
\label{eq:eigenproblem}
- \Delta u(x,\y) = \lambda u(x,\y) \quad (x,\y) \in D, \quad \quad u|_{\dc D} = 0 ,
\end{equation}
where $ \oe D = \oe V \cup \oe Q $ is decomposed into a basic bounded
domain $V\subset \R^{n+1}$ of arbitrary shape and a branch $Q\subset
\R^{n+1}$ of a variable cross-section profile $\Omega(x)\subset \R^n$
(Fig. \ref{fig:domain}b,c):
\begin{equation*}
Q = \{(x,\y)\in \R^{n+1}~:~\y \in \Omega(x),~ 0 < x < a \}.
\end{equation*}
Each cross-section $\Omega(x)$ is a bounded domain which is
parameterized by $x$ from $0$ to $a$.  The boundary of the branch $Q$
is assumed to be piecewise smooth \cite{Grisvard}.  Although weaker
conditions on the boundary could potentially be used, their
justification would require a substantial technical analysis which is
beyond the scope of the paper.

For a fixed $x$, let $\mu_1(x)$ be the first eigenvalue of the problem
\begin{equation}
- \Delta_{\pe} \phi(\y) = \mu_1(x) \phi(\y)  \quad \y \in \Omega(x), \quad \quad \phi|_{\dc \Omega} = 0  ,
\end{equation}
where $\Delta_{\pe}$ is the $n$-dimensional Laplace operator.  We
denote
\begin{equation}
\label{eq:mu}
\mu = \inf\limits_{0<x<a} \mu_1(x) 
\end{equation}
the smallest first eigenvalue among all cross-sections of the branch.
For example, if $\Omega(x) = [0,b]$ (independent of $x$), one has $\mu
= \pi^2/b^2$ and retrieves the example from
Sec. \ref{sec:rectangular}.

Now, we can formulate the main result of the paper.  If the basic
domain $V$ is large enough so that
\begin{equation}
\label{eq:assumpt}
\lambda < \mu ,
\end{equation}
we prove that the squared $L_2$-norm of the eigenfunction $u$ in the
cross-section $\Omega(x_0)$,
\begin{equation}
\label{eq:Ix}
I(x_0) \equiv \int \limits_{\Omega(x_0)} u^2(x_0,\y) d\y ,
\end{equation}
decays exponentially with $x_0$:
\begin{equation}
\label{eq:expon}
I(x_0) \leq I(0) e^{- \beta x_0} \quad (0\leq x_0 < a),
\end{equation}
with the decay rate 
\begin{equation}
\label{eq:beta1}
\beta = \sqrt{2}~ \sqrt{\mu - \lambda}.
\end{equation}
Moreover, if the branch profile $\Omega(x)$ satisfies the condition
\begin{equation}
\label{eq:cond}
(\e_x , \n(x,\y)) \geq 0  \quad \forall (x,\y)\in \dc Q ,
\end{equation}
where $\e_x$ is the unit vector along the $x$ coordinate, and
$\n(x,\y)$ the unit normal vector at the boundary point $(x,\y)$
directed outwards the domain, then the decay rate is improved:
\begin{equation}
\label{eq:beta2}
\beta = 2\sqrt{\mu - \lambda}.
\end{equation}
Qualitatively, the condition (\ref{eq:cond}) means that the branch $Q$
is not increasing.  Note that the example of a rectangular branch from
Sec. \ref{sec:rectangular} shows that the decay rate in
Eq. (\ref{eq:beta2}) is sharp (cannot be improved).

The proof consists in three steps.

(i) First, we derive the inequality
\begin{equation}
\label{eq:Maslov} 
I''(x_0) \geq c\gamma^2 I(x_0)  ,
\end{equation}
where $c = 2$ for arbitrary branch and $c = 4$ for branches satisfying
the condition (\ref{eq:cond}), and $\gamma = \sqrt{\mu - \lambda}$.
This type of inequalities was first established by Maslov for
Schr\"odinger operators in free space \cite{Maslov64,Maslov}.  For
this purpose, we consider the first two derivatives of $I(x_0)$:
\begin{eqnarray}
\label{eq:aux2}
I'(x_0) & = & 2 \int\limits_{\Omega(x_0)} u \frac{\dc u}{\dc x}~ d\y , \\
\label{eq:aux3}
I''(x_0) & = & 2 \int\limits_{\Omega(x_0)} u \frac{\dc^2 u}{\dc x^2}~ d\y 
+ 2 \int\limits_{\Omega(x_0)} \biggl(\frac{\dc u}{\dc x}\biggr)^2 d\y , 
\end{eqnarray}
where the boundary condition $u|_{\dc Q} = 0$ cancels the integrals
over the ``lateral'' boundary of $Q(x_0)$.

The first integral in Eq. (\ref{eq:aux3}) can be estimated as
\begin{equation}
\label{eq:aux8}
\begin{split}
& \int\limits_{\Omega(x_0)} u \frac{\dc^2 u}{\dc x^2} d\y =
\int\limits_{\Omega(x_0)} u \bigl[\Delta u - \Delta_{\pe} u\bigr] ~d\y  \\
& = \int\limits_{\Omega(x_0)}(\nabla_{\pe} u, \nabla_{\pe} u) d\x - 
\lambda \int\limits_{\Omega(x_0)} u^2 d\y  \geq (\mu - \lambda) \int \limits_{\Omega(x_0)} u^2 d\y , \\
\end{split}
\end{equation}
where we used the Friedrichs-Poincar\'e inequality for the section
$\Omega(x_0)$ (see Appendix \ref{sec:Friedrichs}):
\begin{equation}
\label{eq:Poincare}
\int\limits_{\Omega(x_0)} (\nabla_\perp u, \nabla_\perp u) d\y \geq 
\mu_1(x_0) \int\limits_{\Omega(x_0)} u^2 d\y \geq \mu \int \limits_{\Omega(x_0)} u^2 d\y ,
\end{equation}
and $\mu_1(x_0) \geq \mu$ by definition of $\mu$ in Eq. (\ref{eq:mu}).
Since the second term in Eq. (\ref{eq:aux3}) is always positive, one
has
\begin{equation*}
I''(x_0) \geq 2 \int \limits_{\Omega(x_0)} u \frac{\dc^2 u}{\dc x^2}  d\y \geq 2 (\mu - \lambda) \int \limits_{\Omega(x_0)} u^2 d\y,
\end{equation*}
from which follows the inequality (\ref{eq:Maslov}) with $c = 2$.

If the condition (\ref{eq:cond}) is satisfied, a more accurate
estimate of the second term in Eq. (\ref{eq:aux3}) follows from the
Rellich's identity (see Appendix \ref{sec:Rellich}):
\begin{equation}
\label{eq:Rellich}
\begin{split}
& \int\limits_{\Omega(x_0)} \biggl(\frac{\dc u}{\dc x}\biggr)^2 d\y = \int\limits_{\Omega(x_0)} (\nabla_{\pe} u, \nabla_{\pe} u) d\y 
- \lambda \int\limits_{\Omega(x_0)} u^2 d\y  \\
& + \int\limits_{\dc Q(x_0)\backslash \Omega(x_0)} \biggl(\frac{\dc u}{\dc n}\biggr)^2 (\e_x, \n(S)) dS , \\
\end{split}
\end{equation}
where $Q(x_0)$ denotes the ``right'' part of the branch $Q$ delimited
by $\Omega(x_0)$:
\begin{equation}
\label{eq:Qx0}
Q(x_0) = \left\{(x,\y)\in\R^{n+1}~:~ \y \in \Omega(x), ~x_0 < x < a \right\} .
\end{equation}
The condition (\ref{eq:cond}) implies the positivity of the last term
in Eq. (\ref{eq:Rellich}), which can therefore be dropped off in order
to get the following estimate:
\begin{equation*}
\int\limits_{\Omega(x_0)} \biggl(\frac{\dc u}{\dc x}\biggr)^2 d\y  \geq 
\int\limits_{\Omega(x_0)} (\nabla_{\pe} u, \nabla_{\pe} u) d\y - \lambda \int\limits_{\Omega(x_0)} u^2 d\y 
  \geq  (\mu - \lambda) \int\limits_{\Omega(x_0)} u^2 d\y . 
\end{equation*}
Combining this result with (\ref{eq:aux8}), one gets the inequality
(\ref{eq:Maslov}) with $c = 4$.

(ii) Second, we prove the following relations
\begin{equation}
\label{eq:relations} 
I(a) = 0, \quad I'(a) = 0, \quad I(x_0) \ne 0, \quad  I'(x_0) < 0  
\end{equation}
for $0\leq x_0 < a$.  In fact, taking into account Eq. (\ref{eq:aux2})
and applying the Green's formula in subdomain $Q(x_0)$, one obtains
\begin{equation}
\label{eq:aux1}
I'(x_0) = - 2 \int\limits_{Q(x_0)} u \Delta u ~ dx d\y - 2 \int \limits_{Q(x_0)} (\nabla u, \nabla u) dx d\y   ,
\end{equation}
where the boundary condition $u|_{\partial Q(x_0)\backslash
\Omega(x_0)} = 0$ canceled boundary integrals.  The second term can be
estimated by using again the Friedrichs-Poincar\'e inequality
(\ref{eq:Poincare}):
\begin{equation}
\label{eq:Poincare2}
\begin{split}
& \int\limits_{Q(x_0)} \hspace*{-1mm} (\nabla u, \nabla u) dx d\y = 
\hspace*{-1mm} \int\limits_{Q(x_0)} \biggl[\biggl(\frac{\partial u}{\partial x}\biggr)^2 + (\nabla_\perp u, \nabla_\perp u) \biggr] dx d\y  \\
& \geq \int\limits_{x_0}^a dx \int\limits_{\Omega(x)} (\nabla_\perp u, \nabla_\perp u)  d\y
\geq \mu \int\limits_{x_0}^a dx \int \limits_{\Omega(x)} u^2 dx d\y 
 = \mu \int \limits_{Q(x_0)} u^2 dx d\y . \\
\end{split}
\end{equation}
The equation $\Delta u = - \lambda u$ yields then
\begin{equation*}
\begin{split}
- I'(x_0) & = - 2 \lambda \int \limits_{Q(x_0)} u^2 dx d\y + 2 \int \limits_{Q(x_0)} (\nabla u, \nabla u) dx d\y \\
& \geq 2(\mu - \lambda) \int \limits_{Q(x_0)} u^2 dx d\y \geq 0 ,
\end{split}
\end{equation*}
so that $I(x_0)$ monotonously decays.  We emphasize that there is no
request on a monotonous decrease of any kind for the cross-section
$\Omega(x)$.

Sending $x_0$ to $a$ in Eq. (\ref{eq:aux1}), one gets $I'(a) = 0$.

Finally, we show that $I(x_0) \ne 0$ for $0 \leq x_0 < a$.  Indeed, if
there would exist $x_0$ such that $I(x_0) = 0$, then the restriction
of $u$ to the subdomain $Q(x_0)$ is a solution of the eigenvalue
problem in $Q(x_0)$:
\begin{equation*}
 - \Delta u = \lambda u \quad (x,\y) \in Q(x_0), \quad \quad u|_{\dc Q(x_0)} = 0 .
\end{equation*}
Multiplying this equation by $ u $ and integrating over the domain $
Q(x_0) $ lead to
\begin{equation*}
\lambda \int \limits_{Q(x_0)} u^2 dx d\y = \int \limits_{Q(x_0)} (\nabla u, \nabla u) dx d\y .
\end{equation*}
On the other hand, the Friedrichs-Poincar\'e inequality
(\ref{eq:Poincare2}) yields
\begin{equation*}
\int\limits_{Q(x_0)} (\nabla u,\nabla u) dx d\y \geq \mu \int \limits_{Q(x_0)} u^2 dx d\y , 
\end{equation*}
from which $\lambda \geq \mu$, in contradiction to the condition
(\ref{eq:assumpt}).

(iii) Third, we establish the exponential decay (\ref{eq:expon})
following the Maslov's method \cite{Maslov64,Maslov}.  The
multiplication of the inequality (\ref{eq:Maslov}) by $I'(x_0)$ yields
\begin{equation*}
((I')^2)' \leq c \gamma^2 (I^2)' .
\end{equation*}
After integrating from $x_0$ to $a$, one gets 
\begin{equation*}
- (I'(x_0))^2 \leq - c\gamma^2 I^2(x_0) ,
\end{equation*} 
where $I(a) = I'(a) = 0$ from Eq. (\ref{eq:relations}).  Taking in
account that $I' < 0$, one deduces
\begin{equation*}
- I'(x_0) \geq \sqrt{c} \gamma I(x_0) .
\end{equation*}
Dividing by $I(x_0)$ and integrating from $0$ to $x_0$ lead to
Eq. (\ref{eq:expon}) that completes our proof.

As in Sec. \ref{sec:rectangular}, no information about the basic
domain $V$ was used so that the Dirichlet boundary condition on $\dc
V$ can be replaced by arbitrary boundary condition under which the
Laplace operator in $D$ remains self-adjoint.

\section{ Discussion }
\label{sec:discussion}

The main result of Sec. \ref{sec:general} states that if an eigenvalue
$\lambda$ is below the smallest eigenvalue $\mu$ in all cross-sections
$\Omega(x)$, the associated eigenfunction $u$ decays exponentially in
the branch $Q$.  This condition on the eigenvalue $\lambda$ in $D$ can
be replaced by another condition
\begin{equation}
\label{eq:assumpt1}
\kappa < \mu
\end{equation}
on the eigenvalue $\kappa$ in the basic domain $V$:
\begin{equation}
\label{2} 
 - \Delta \phi = \kappa \phi, \quad (x,\y) \in V, \quad \quad \phi|_{\dc V} = 0 .
\end{equation}
According to the Rayleigh's principle, the inequality
(\ref{eq:assumpt1}) is sufficient for the existence of the eigenvalue
$\lambda$ lying below $\mu$ (see Appendix \ref{sec:rayleigh}).
Although the condition (\ref{eq:assumpt1}) is {\it stronger} than
(\ref{eq:assumpt}), it may be preferred for studying different
branches $Q$ attached to the same domain $V$ (in which case, the
eigenvalue $\kappa$ has to be computed only once).

\begin{figure}
\begin{center}
\includegraphics[width=100mm]{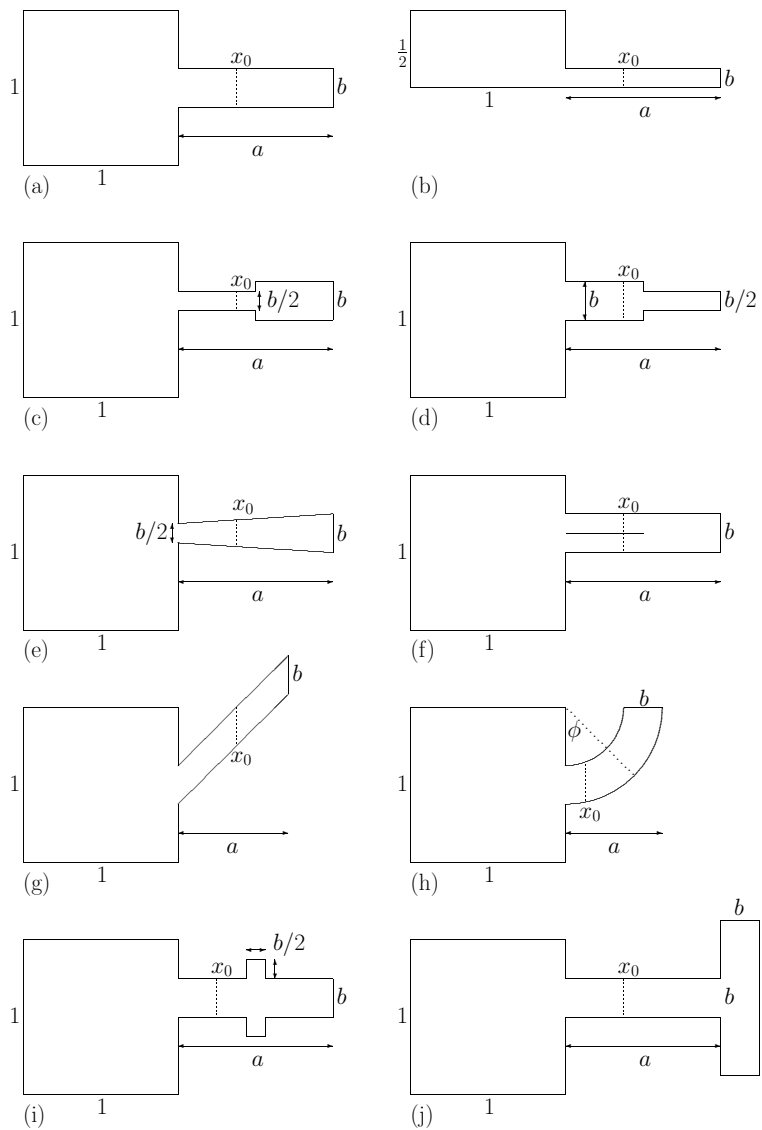}
\end{center}
\caption{
The unit square (basic domain $V$) and several shapes of the branch
$Q$: {\bf (a)} rectangular branch; {\bf (b)} half of the domain 'a';
{\bf (c)} narrow-than-wide channel; {\bf (d)} wide-than-narrow
channel; {\bf (e)} increasing branch with the width linearly changing
from $b/2$ to $b$; {\bf (f)} branch with a partial cut at the middle;
{\bf (g)} tilted rectangular branch; {\bf (h)} circular branch; {\bf
(i)} branch with a small broadening in the middle; {\bf (j)}
bifurcating branch.  We set $a = 1$ and $b = 1/4$ in all cases, except
'g' and 'h', for which $a = 1/\sqrt{2}$ and $a = 5/8$, respectively.
For shapes 'c', 'd', 'e', 'f', 'i' and 'j', the branch is up shifted
by $1/8$ in order to break the reflection symmetry (for cases 'g' and
'h', there is no shift because the branch itself has no reflection
symmetry). }
\label{fig:domains}
\end{figure}

Looking at the derivation of the inequality (\ref{eq:expon}), many
questions naturally appear: How accurate the exponential estimate is?
Is the condition (\ref{eq:cond}) necessary for getting the sharp decay
rate according to Eq. (\ref{eq:beta2})?  How do the eigenfunctions
with $\lambda$ larger than $\mu$ behave?  Is the exponential decay
applicable for other boundary conditions?  In the next subsection, we
address these questions through numerical simulations in planar
domains.

\begin{table}
\begin{center}
\begin{tabular}{| c | c | c | c | c | c | c | c | c | c |}  \hline
$n$& 2a & 2c & 2d & 2e & 2f & 2g & 2h & 2i & 2j \\  \hline
 1 & 19.33 & 19.66 & 19.39 & 19.64 & 19.58 & 19.39 & 19.33 & 19.39 & 19.39 \\ 
 2 & 47.53 & 48.95 & 47.58 & 48.94 & 48.59 & 47.86 & 47.54 & 47.58 & 47.58 \\ 
 3 & 49.32 & 49.35 & 49.33 & 49.35 & 49.33 & 49.32 & 49.32 & 49.33 & 49.33 \\ 
 4 & 78.83 & 78.75 & 77.85 & 78.90 & 78.54 & 78.85 & 78.84 & 77.85 & 77.85 \\ 
 5 & 93.12 & 97.87 & 94.41 & 97.69 & 97.09 & 94.48 & 93.12 & 94.40 & 94.41 \\ 
 6 & 98.70 & 98.71 & 98.67 & 98.71 & 98.66 & 98.71 & 98.71 & 98.67 & 98.67 \\ 
 7 & 126.1 & 127.8 & 125.2 & 127.9 & 127.3 & 126.9 & 126.1 & 125.2 & 125.2 \\ 
 8 & 128.0 & 128.3 & 128.0 & 128.3 & 128.0 & 128.1 & 128.0 & 128.0 & 127.7 \\ 
 9 & 151.7 & 166.1 & 154.1 & 165.8 & 164.6 & 158.5 & 151.6 & 150.4 & 128.0 \\ 
10 & 167.7 & 167.8 & 167.8 & 167.8 & 167.8 & 167.7 & 167.7 & 156.8 & 153.6 \\ 
11 & 167.8 & 177.5 & 176.5 & 177.1 & 177.0 & 175.8 & 171.8 & 167.8 & 167.8 \\ 
12 & 175.3 & 193.9 & 182.3 & 196.9 & 183.4 & 196.5 & 176.1 & 176.7 & 167.8 \\ 
13 & 191.7 & 196.1 & 196.1 & 197.5 & 195.3 & 197.4 & 196.4 & 185.3 & 176.7 \\ 
14 & 196.4 & 197.5 & 197.4 & 239.6 & 197.4 & 229.6 & 197.4 & 197.1 & 185.4 \\ 
15 & 197.4 & 245.1 & 229.1 & 244.4 & 244.1 & 245.9 & 204.5 & 197.4 & 195.2 \\ 
16 & 218.8 & 246.8 & 246.0 & 246.8 & 246.2 & 250.7 & 235.5 & 218.5 & 195.5 \\ 
17 & 245.7 & 254.3 & 249.2 & 254.1 & 252.4 & 256.7 & 245.8 & 242.8 & 197.4 \\ 
18 & 246.7 & 256.7 & 256.6 & 256.7 & 256.6 & 284.6 & 250.5 & 246.1 & 214.8 \\ 
19 & 252.5 & 284.7 & 269.6 & 285.8 & 259.3 & 285.5 & 256.7 & 250.6 & 229.0 \\ 
20 & 256.6 & 286.3 & 285.9 & 286.3 & 283.2 & 304.1 & 278.7 & 256.6 & 245.9 \\  \hline
\end{tabular}
\end{center}
\caption{
First 20 eigenvalues of the Laplace operator in domains shown on
Fig. \ref{fig:domains}. }
\label{tab:lambda}
\end{table}

\subsection{ Two-dimensional branches }

\begin{figure}
\begin{center}
\includegraphics[width=85mm]{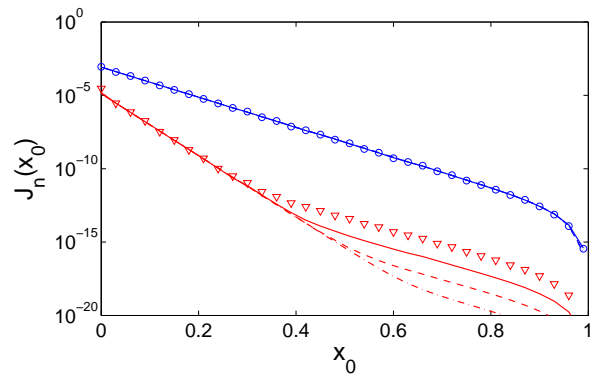}
\end{center}
\caption{
(Color online) Computation of $J_n(x_0)$ at different levels $k$ of
mesh refinement of the rectangular branch on Fig. \ref{fig:domains}a:
$k=3$ (symbols, 6848 triangles in the mesh), $k=4$ (solid lines, 27392
triangles), $k=5$ (dashed lines, 109568 triangles) and $k=6$
(dash-dotted lines, 438272 triangles).  For $n=1$, all these curves
fall onto each other, confirming the accurate computation which is
independent of the mesh size.  In turn, the curves for $n=3$ coincide
only for small $x_0$ but deviate from each other for larger $x_0$.
The higher $k$, the closer the curve to the expected exponential decay
(Fig. \ref{fig:rectangle_J}). }
\label{fig:J_check}
\end{figure}

In order to answer these and some other questions, we consider several
planar bounded domains $D$ which are all composed of the unit square
$V$ as the basic domain and a branch $Q$ of different shapes
(Fig. \ref{fig:domains}).  The eigenvalue problem
(\ref{eq:eigenproblem}) in these domains with Dirichlet boundary
condition was solved numerically by Matlab PDEtools.  Once the
eigenfunctions and eigenvalues are found, we approximate the squared
$L_2$-norm of the eigenfunction $u_n$ in the subregion $Q(x_0) =
\{(x,y)\in Q~:~ x_0 < x < a\}$ of the branch $Q$ as
\begin{equation}
\label{eq:J_num}
J_n(x_0) \equiv \int\limits_{Q(x_0)} u_n^2(x,y) dx dy \simeq \sum\limits_{T} \frac{S(T)}{3} \sum\limits_{j=1}^3 u_n^2(x^T_j,y^T_j),
\end{equation}
where the sum runs over all triangles $T$ of the mesh, $S(T)$ being
the area of the triangle $T$, and $\{(x^T_j,y^T_j)\}_{j=1,2,3}$ its
three vertices.  Since the eigenfunctions are analytic inside the
domain, the error of the above approximation is mainly determined by
the areas of triangles.  In order to check whether the results are
accurate or not, we computed the function $J_n(x_0)$ at different
levels $k$ of mesh refinement (once the initial triangular mesh is
generated by Matlab, each level of refinement consists in dividing
each triangle of the mesh into four triangles of the same shape).
Figure \ref{fig:J_check} shows the resulting curves for the
rectangular branch (Fig. \ref{fig:domains}a).  For the first
eigenfunction, all the curves fall onto each other, i.e. $J_1(x_0)$ is
independent of $k$, as it should be.  In turn, the curves for $n=3$
coincide only for small $x_0$ but deviate from each other for larger
$x_0$.  The higher $k$, the closer the curve to the expected
exponential decay.  This means that even 6 levels of mesh refinement
(i.e., a mesh with 438272 triangles) is not enough for an accurate
computation of the integral $J_3(x_0)$.  Among the 20 first
eigenfunctions, similar deviations were observed for $n = 3, 4, 8, 10,
14, 15, 17$.  The specific behavior of these eigenfunctions seems to
be related to their reflection symmetry.  Although several
improvements could be performed [e.g., higher-order integration
schemes instead of Eq. (\ref{eq:J_num})], this is too technical and
beyond the scope of the paper because our numerical computations aim
only at illustrating the theoretical estimates.  After all, the
deviations become distinguishable only in the region of $x_0$ for
which $J_n(x_0)$ is negligible.  For all data sets discussed below, we
checked the accuracy by performing computations with different $k$ and
presented only the reliable data with $k = 5$ (such meshes contain
between $100000$ and $170000$ triangles, except for the case on
Fig. \ref{fig:domains}f with 671744 triangles).

The exponential decay (\ref{eq:expon}) for $I(x_0)$ implies that of
$J_n(x_0)$:
\begin{equation}
\label{eq:J_ineq}
J_n(x_0) = \int\limits_{x_0}^a dx~ I_n(x) \leq \int\limits_{x_0}^a dx~ I_n(0) e^{-2\gamma_n x} \leq 
\frac{I_n(0)}{2\gamma_n} e^{-2\gamma_n x_0} ,
\end{equation}
where 
\begin{equation*}
\gamma_n = \sqrt{\mu - \lambda_n} ,
\end{equation*}
and we take $c = 4$ even if the sufficient condition (\ref{eq:cond})
is not satisfied.  In what follows, we will check numerically the
stronger inequality
\begin{equation}
\label{eq:J_ineq3}
J_n(x_0) \leq J_n(0) e^{-2\gamma_n x_0} ,
\end{equation}
from which (\ref{eq:J_ineq}) follows, because
\begin{equation*}
J_n(0) = \int\limits_0^a dx I_n(x) \leq I_n(0) \int\limits_0^a dx e^{-2\gamma_n x} \leq \frac{I_n(0)}{2\gamma_n} .
\end{equation*}

It is worth stressing that the inequality (\ref{eq:J_ineq3}) for the
squared $L_2$-norm $J_n(x_0)$ in the subregion $Q(x_0)$ is a {\it
weaker} result than the inequality (\ref{eq:expon}) for the squared
$L_2$-norm $I_n(x_0)$ in the cross-section $\Omega(x_0)$.  However,
the analysis of $I_n(x_0)$ would require an accurate computation of
the projection of an eigenfunction $u_n$, which was computed on a
triangular mesh in $D$, onto the cross-section $\Omega(x_0)$.  Since
the resulting $I_n(x_0)$ would be less accurate than $J_n(x_0)$, we
focus on the latter quantity.

\begin{figure}
\begin{center}
\includegraphics[width=85mm]{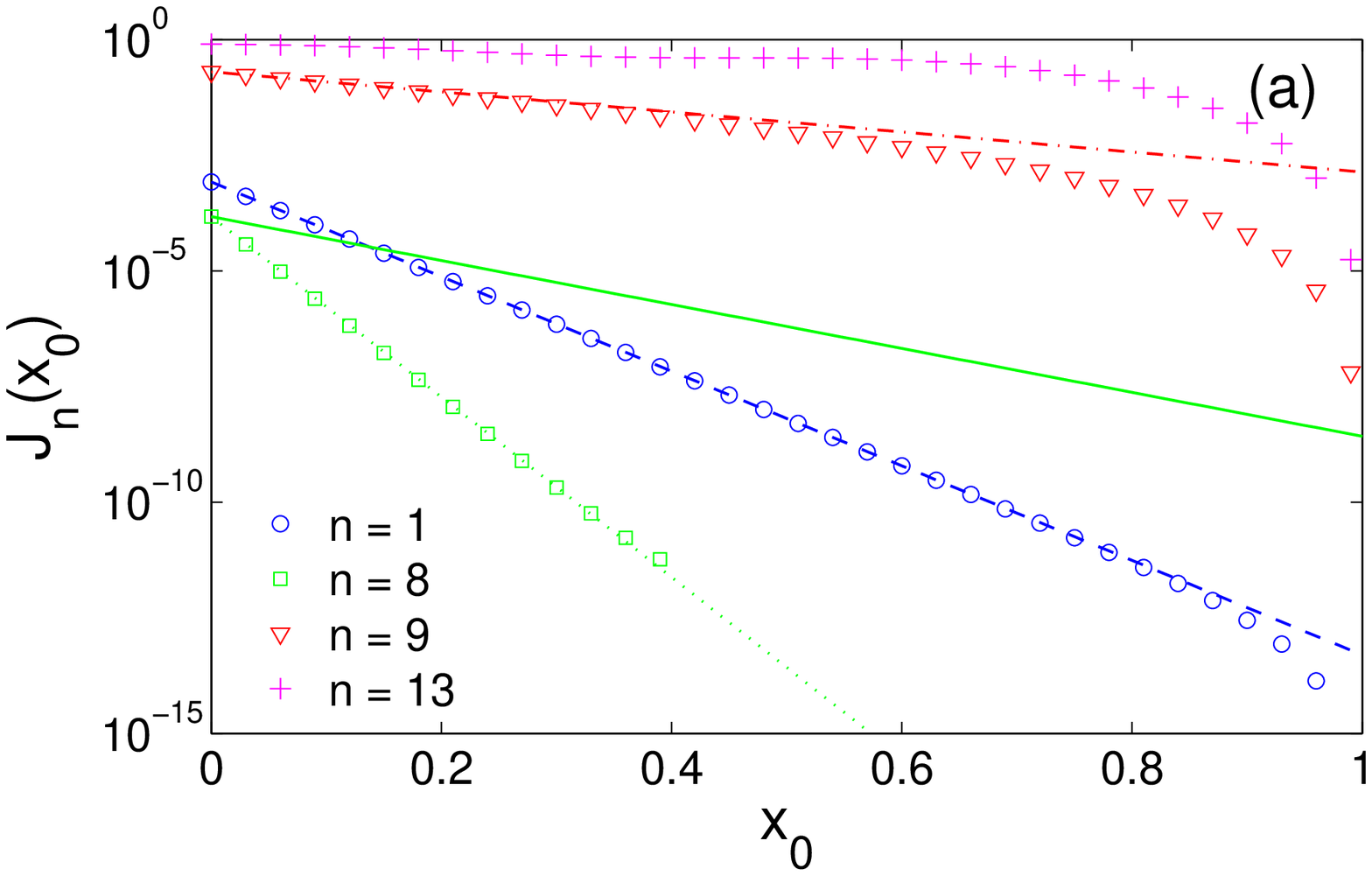}
\includegraphics[width=75mm]{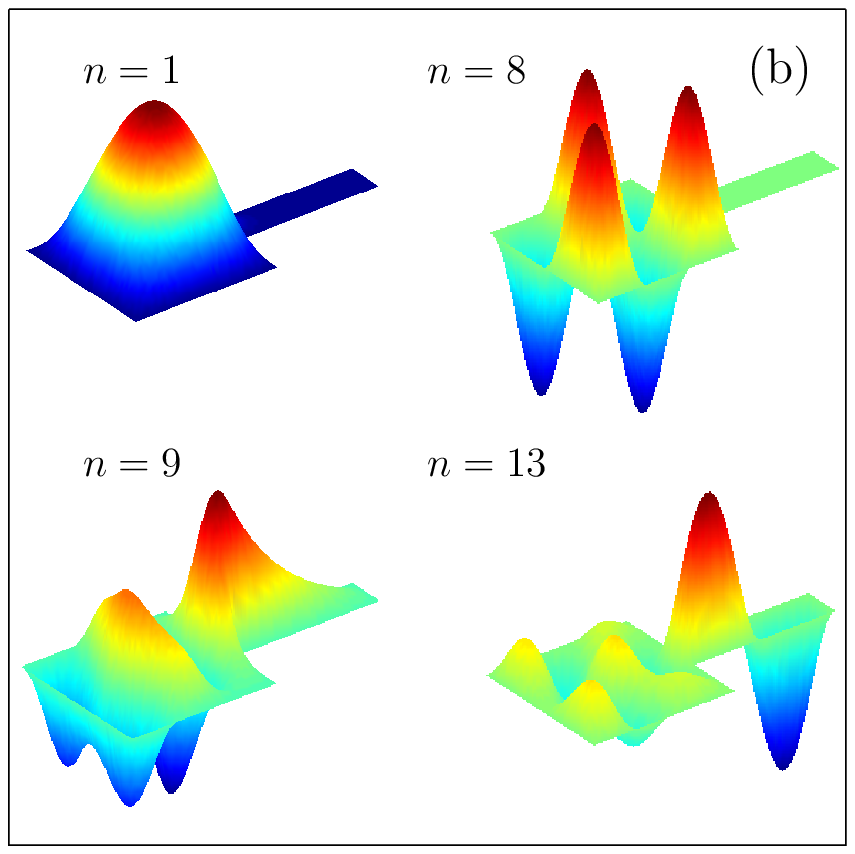}
\end{center}
\caption{
(Color online) The squared $L_2$-norm, $J_n(x_0)$, of four
eigenfunctions with $n = 1, 8, 9, 13$ (symbols) for the rectangular
branch on Fig. \ref{fig:domains}a.  The estimate (\ref{eq:J_ineq3})
with $\mu = \pi^2/(1/4)^2$ is plotted by dashed ($n = 1$), solid ($n =
8$) and dash-dotted ($n = 9$) lines.  The estimate with $\mu =
\pi^2/(1/8)^2$ is shown for $n = 8$ by dotted line.  }
\label{fig:rectangle_J}
\end{figure}

\subsubsection*{ Rectangular branch }

We start with a rectangular branch of width $b = 1/4$
(Fig. \ref{fig:domains}a), for which $\mu = \pi^2/(1/4)^2 \simeq
157.91...$, and there are 9 eigenvalues $\lambda_n$ below $\mu$ (the
first 20 eigenvalues are listed in Table \ref{tab:lambda}).  Figure
\ref{fig:rectangle_J}a shows $J_n(x_0)$ for four eigenfunctions with $n
= 1, 8, 9, 13$.  These four eigenfunctions are chosen to illustrate
different possibilities.  For the eigenmodes with $n = 1,2,5,7,9$
(illustrated by $n = 1,9$), the estimate is very accurate.  In this
case, the decay rate $2\gamma_n$ is sharp and cannot be improved.
There is another group of eigenfunctions with $n = 3,4,6,8$
(illustrated by $n=8$), for which $J_n(x_0)$ is significantly smaller
than the estimate.  For $x_0 < 0.4$, $J_8(x_0)$ decays as
$J_8(0)\exp[-2\gamma'_8 x_0]$, where $2\gamma'_8 = 2\sqrt{4\mu -
\lambda_8}$ is the improved decay rate (for larger $x_0$, the
computation is inaccurate as explained earlier and its result is not
shown).  In order to understand this behavior, one can inspect the
shape of the eigenfunction $u_8(x,y)$ shown on
Fig. \ref{fig:rectangle_J}b.  This function is anti-symmetric with
respect to the horizontal line which splits the domain $D$ into two
symmetric subdomains.  As a consequence, $u_8(x,y)$ is $0$ along this
line and it is thus a solution of the Dirichlet eigenvalue problem for
each subdomain (Fig. \ref{fig:domains}b).  The width of the branch in
each subdomain is twice smaller so that one can apply the general
estimate with $\mu' = 4\mu$.  This is a special feature of all
symmetric domains.

If the branch was shifted upwards or downwards, the reflection
symmetry would be broken, and the decay rate $2\gamma_n'$ would not be
applicable any more.  However, if the shift is small, one may still
expect a faster exponential decrease with the decay rate between
$2\gamma_n$ and $2\gamma_n'$.  This example shows that the estimate
(\ref{eq:expon}) may not be sharp for certain eigenfunctions.  At the
same time, it cannot be improved in general, as illustrated by the
eigenfunctions with $n = 1,9$.

The last curve shown on Fig. \ref{fig:rectangle_J}a by pluses,
corresponds to $n = 13$, for which $\lambda_{13} > \mu$, and the
exponentially decaying estimate is not applicable.  One can see that
the function $J_{13}(x_0)$ slowly varies along the branch.  This
behavior is also expected from its shape shown on
Fig. \ref{fig:rectangle_J}b.

\subsubsection*{ Narrow/wide and wide/narrow branches }

\begin{figure}
\begin{center}
\includegraphics[width=85mm]{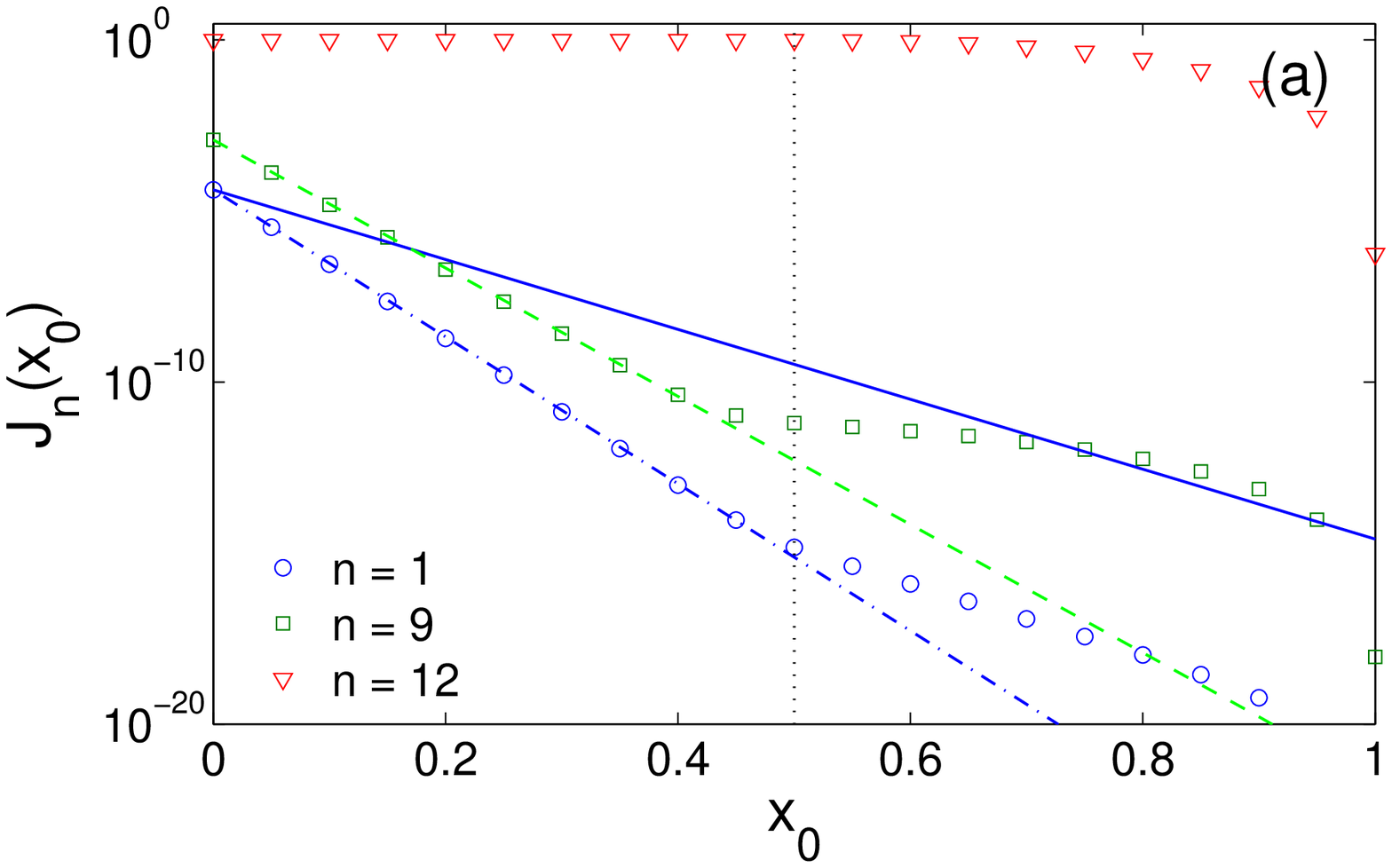}
\includegraphics[width=75mm]{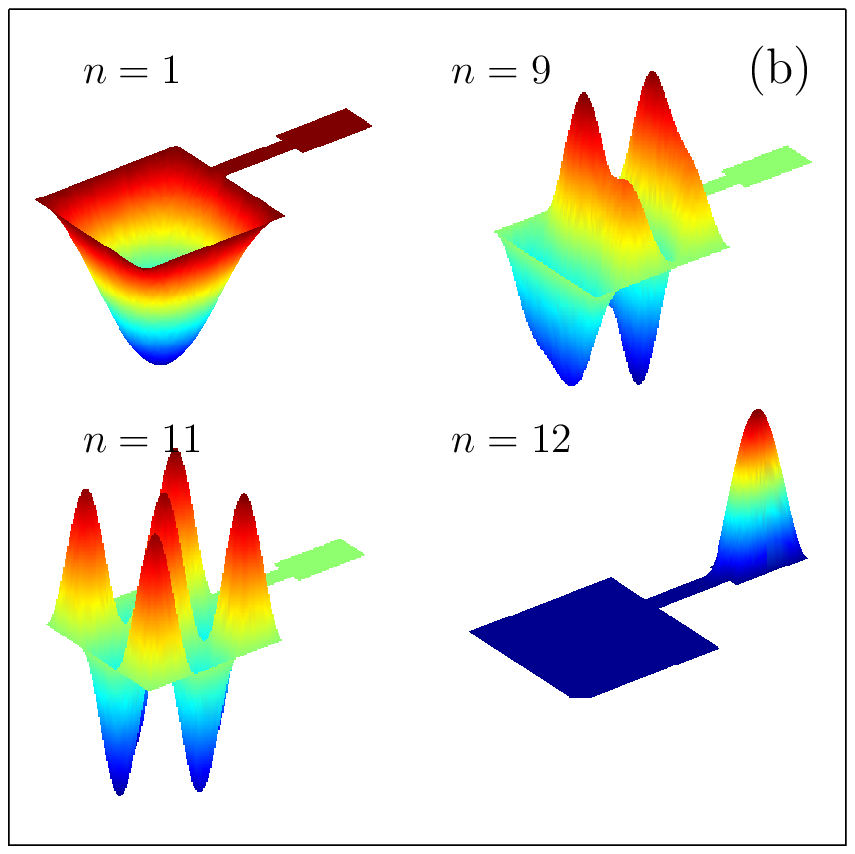}
\end{center}
\caption{
(Color online) The squared $L_2$-norm, $J_n(x_0)$, of three
eigenfunctions with $n = 1, 9, 12$ (symbols) for the narrow-than-wide
branch on Fig. \ref{fig:domains}c.  The estimate (\ref{eq:J_ineq3})
with $\mu = \pi^2/(1/4)^2$ is plotted for $n=1$ by solid line.  The
hypothetical estimate with $\mu = \pi^2/(1/8)^2$ is shown by
dash-dotted ($n = 1$) and dashed ($n = 9$) lines.  The vertical dotted
line indicates the connection between two parts of the branch. }
\label{fig:J_narrow_shifted}
\end{figure}

In two dimensions, any cross-section $\Omega(x)$ of a branch is a
union of intervals.  If $\ell(x)$ is the length of the largest
interval in $\Omega(x)$ then the first eigenvalue $\mu(x)$ in
$\Omega(x)$ is simply $\pi^2/\ell(x)^2$.  Whatever the shape of the
branch is, the bound $\mu$ for the exponential decay is then set by
the length of the largest cross-section $b = \max\limits_{0<x<a}
\ell(x)$ according to $\mu = \pi^2/b^2$.  At first thought, this
statement can sound counter-intuitive because one could expect that
the asymptotic behavior would be determined by the smallest
cross-section.  In order to clarify this point, we consider an example
on Fig. \ref{fig:domains}c for which $\mu = \pi^2/(1/4)^2$.  Although
the narrow channel of width $b/2$ strongly attenuates the amplitude of
the eigenfunction, it does not imply the exponential decay with a
hypothetical rate $2\sqrt{4\mu -\lambda}$ along the next wider channel
(Fig. \ref{fig:J_narrow_shifted}).  For instance, the function
$J_1(x_0)$ exponentially decays with the rate $2\sqrt{4\mu - \lambda}$
(dashed line) up to $x_0 \approx 0.45$.  However, this decay is slowed
down for $x \geq 0.45$.  The theoretical estimate with the decay rate
$2\sqrt{\mu - \lambda}$ (solid line) is of course applicable for all
$x_0$ but it is not sharp.

\begin{figure}
\begin{center}
\includegraphics[width=85mm]{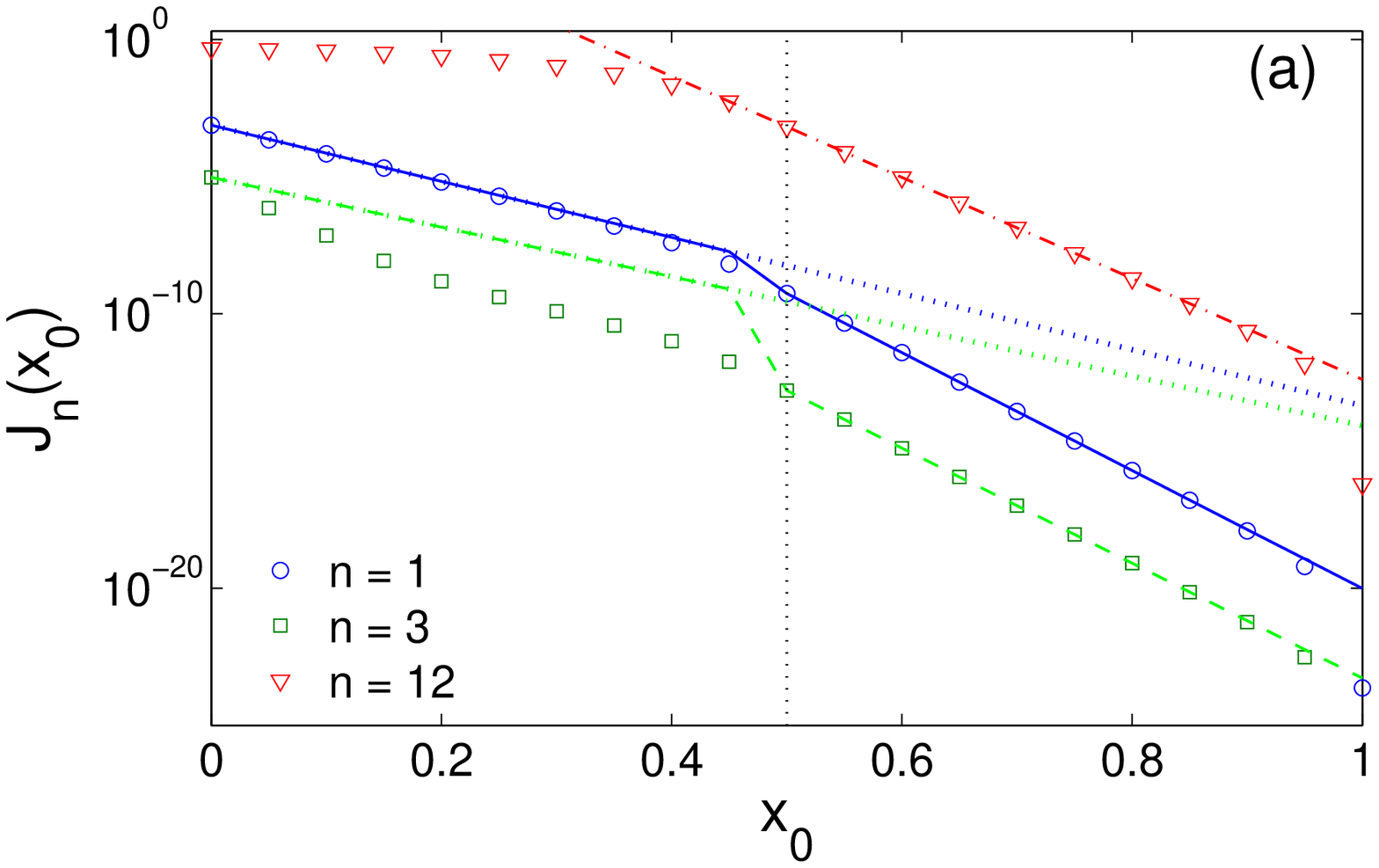}
\includegraphics[width=75mm]{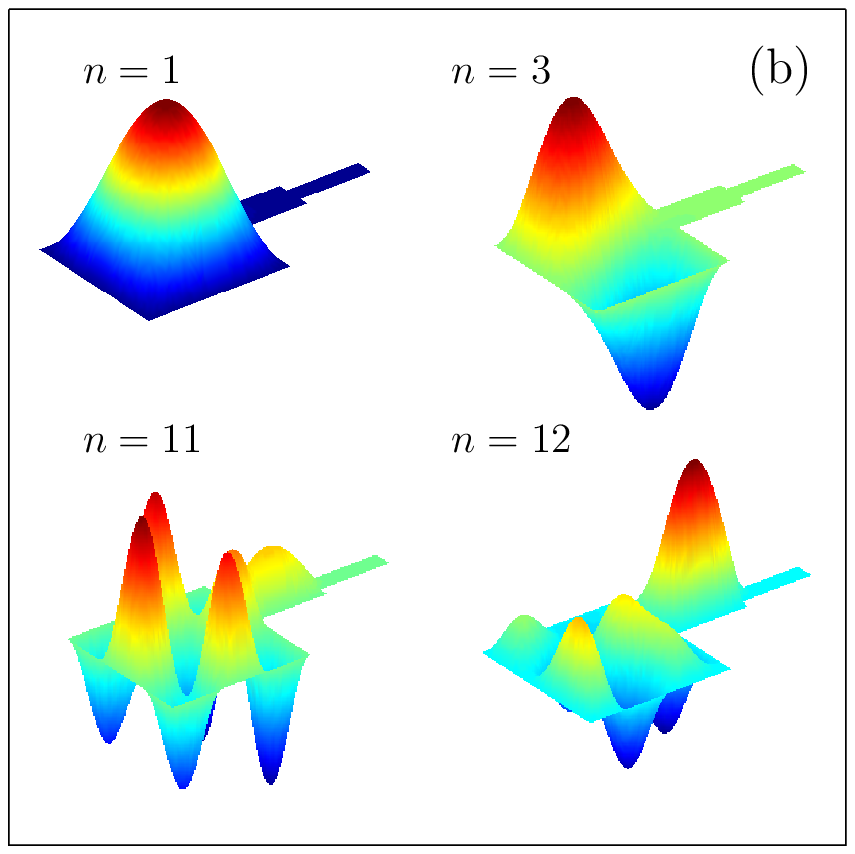}
\end{center}
\caption{
(Color online) The squared $L_2$-norm, $J_n(x_0)$, of three
eigenfunctions with $n = 1, 3, 12$ (symbols) for the wide-than-narrow
branch on Fig. \ref{fig:domains}d.  The estimate (\ref{eq:J_ineq3})
with $\mu = \pi^2/(1/4)^2$ is plotted by dotted lines for $n=1$ and
$n=3$.  The combined estimate (for the wide part with $\mu =
\pi^2/(1/4)^2$ and for the narrow part with $\mu = \pi^2/(1/8)^2$) is
shown by solid ($n = 1$), dashed ($n = 3$) and dash-dotted ($n = 12$)
lines.  The vertical dotted line indicates the connection between two
parts of the branch. }
\label{fig:J_wide_shifted}
\end{figure}

If the wide channel is placed first (as shown on
Fig. \ref{fig:domains}d), the situation is different.  The decay rate
$2\sqrt{\mu-\lambda}$, which is set by the largest cross-section
length $b = 1/4$, can be used in the wide part of the branch.
Although this result is also applicable in the narrow part, the
estimate here can be improved.  In fact, if one considers the unit
square and the wide branch as a basic domain, the decay rate for the
narrow branch, $2\sqrt{4\mu-\lambda}$, is the set by its width $b/2 =
1/8$ (Fig. \ref{fig:J_wide_shifted}).  More generally, if the branch
is a union of several branches with progressively decreasing widths,
one can combine the exponential estimates with progressively
increasing decay rates.

\subsubsection*{ Increasing branch }

\begin{figure}
\begin{center}
\includegraphics[width=85mm]{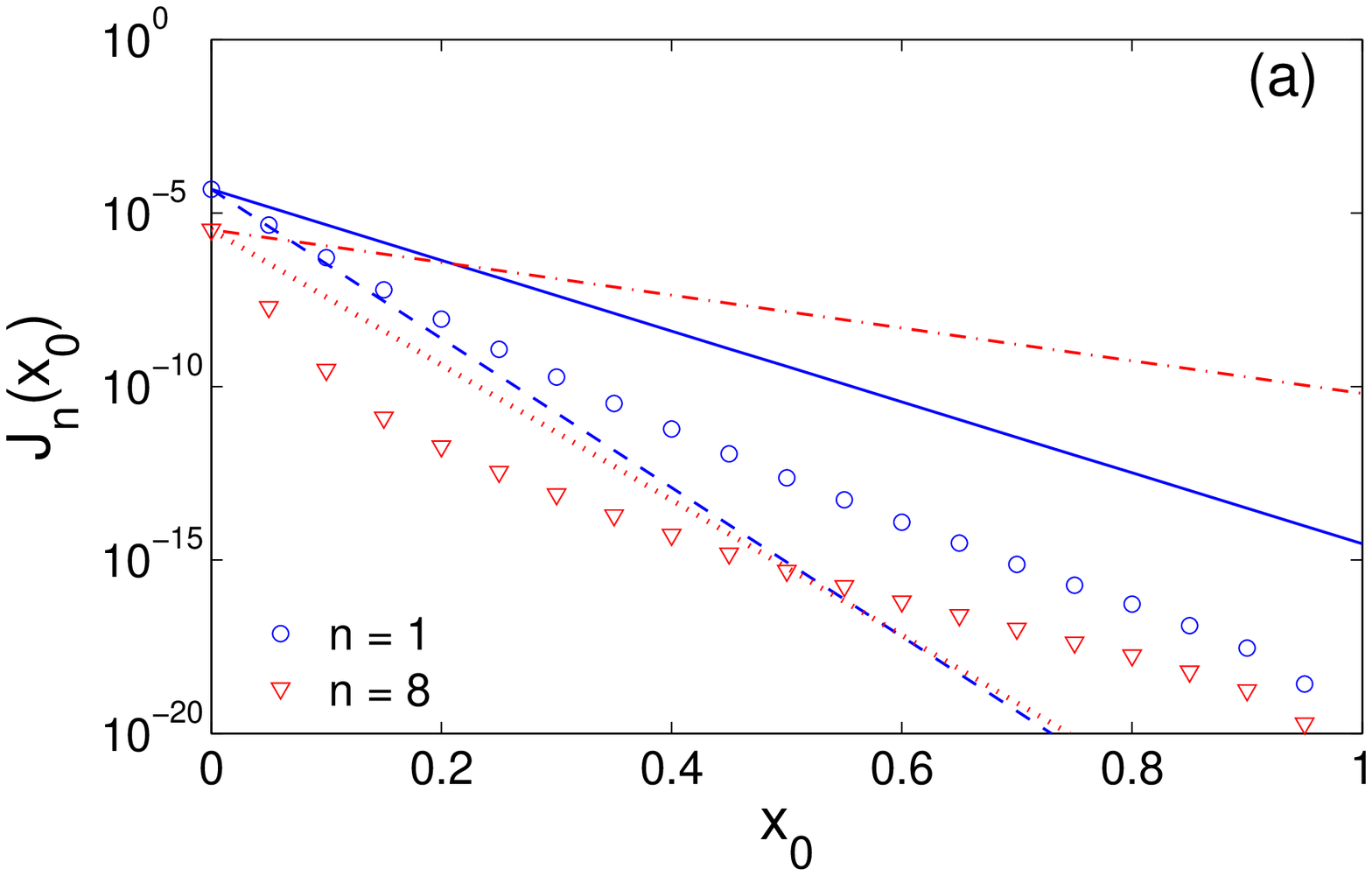}
\includegraphics[width=75mm]{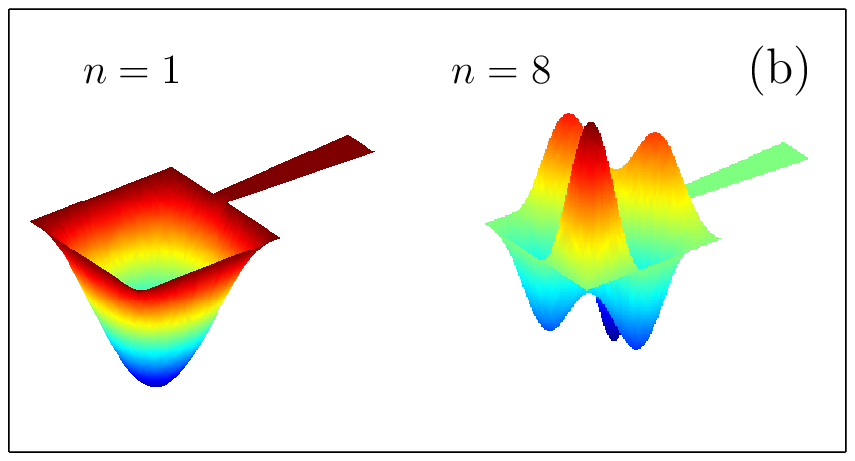}
\end{center}
\caption{
The squared $L_2$-norm, $J_n(x_0)$, of two eigenfunctions with $n = 1,
8$ (symbols) for the increasing branch on Fig. \ref{fig:domains}e.
The estimate (\ref{eq:J_ineq3}) with $\mu = \pi^2/(1/4)^2$ is plotted
by solid ($n=1$) and dash-dotted ($n=8$) lines.  The hypothetical
estimate with $\mu = \pi^2/(1/8)^2$ is shown by dashed ($n = 1$) and
dotted ($n = 8$) lines.  }
\label{fig:J_increasing}
\end{figure}

In the previous example on Fig. \ref{fig:domains}c, the supplementary
condition (\ref{eq:cond}) was not satisfied on a part of the branch
boundary.  Nevertheless, the numerical computation confirmed the sharp
decay rate (\ref{eq:beta2}).  In order to check the relevance of the
condition (\ref{eq:cond}), we consider a linearly increasing branch
shown on Fig. \ref{fig:domains}e.  Although the condition
(\ref{eq:cond}) is not satisfied at any point, the sharp decay rate
(\ref{eq:beta2}) is again applicable, as shown on
Fig. \ref{fig:J_increasing}.  In a future work, one may try to relax
the condition (\ref{eq:cond}) or to provide counter-examples showing
its relevance.

\subsubsection*{ Branch with a cut }

A small variation in the shape of the domain is known to result in
small changes in eigenvalues of the Laplace operator, at least at the
beginning of the spectrum (see Table \ref{tab:lambda}).  In turn, the
eigenfunctions may be very sensitive to any perturbation of the
domain.  Bearing this sensitivity in mind, one can check the
robustness of the exponential decay.  We consider a horizontal cut at
the middle of the rectangular branch (Fig. \ref{fig:domains}f).  If
the cut went along the whole branch, it would be equivalent to two
separate rectangular branches of width $b/2$.  In this case, the
theoretical estimate could be applied individually to each branch, and
the value $\mu = \pi^2/(b/2)^2$ would be 4 times larger than for the
original rectangular branch of width $b$.  For a partial cut, whatever
its length is, the theoretical decay rate is again determined by $\mu
= \pi^2/(1/4)^2$ as for the rectangular branch.  At the same time, it
is clear that the cut results in a stronger ``expelling'' of
eigenfunctions from the branch.

\begin{figure}
\begin{center}
\includegraphics[width=85mm]{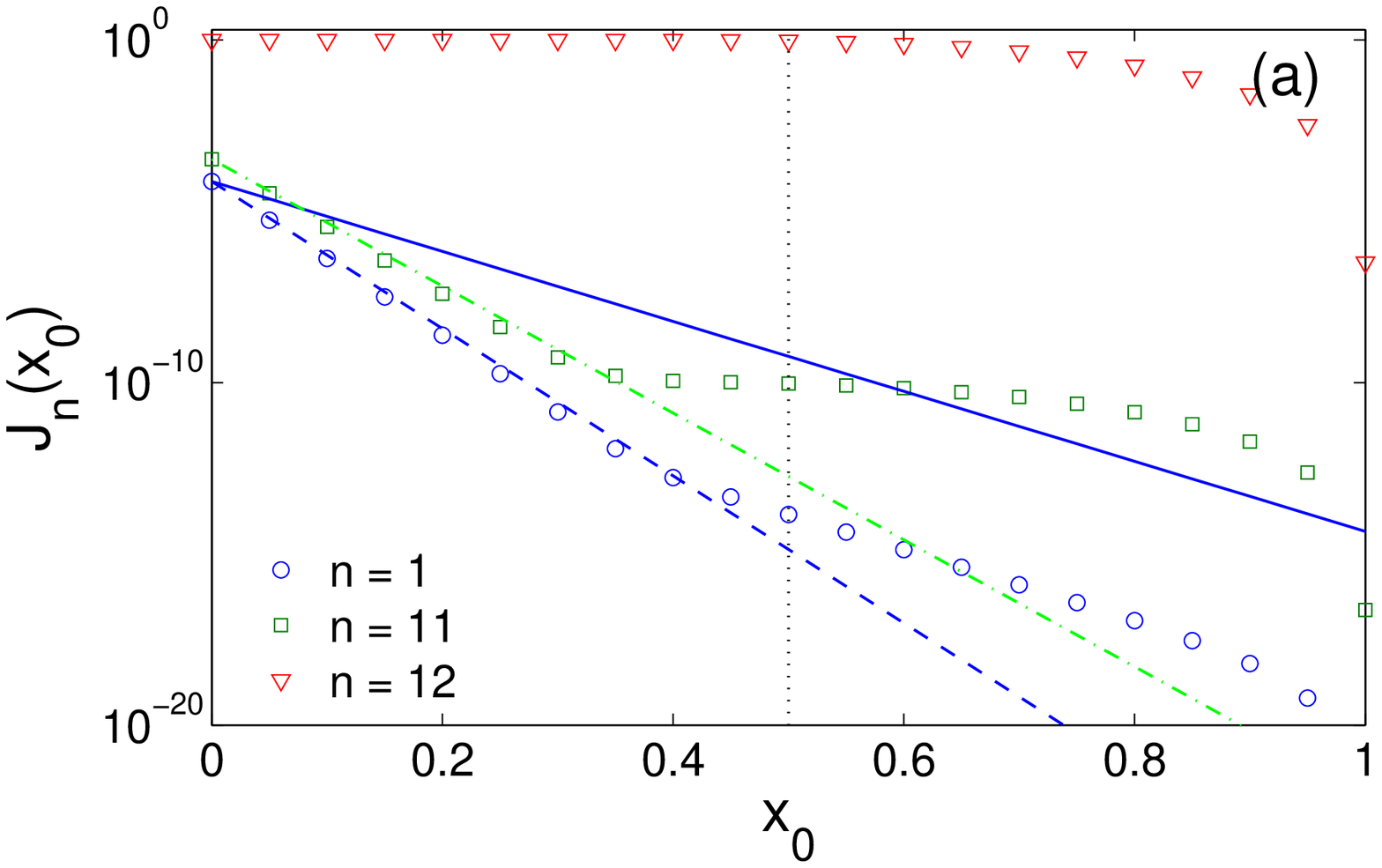}
\includegraphics[width=75mm]{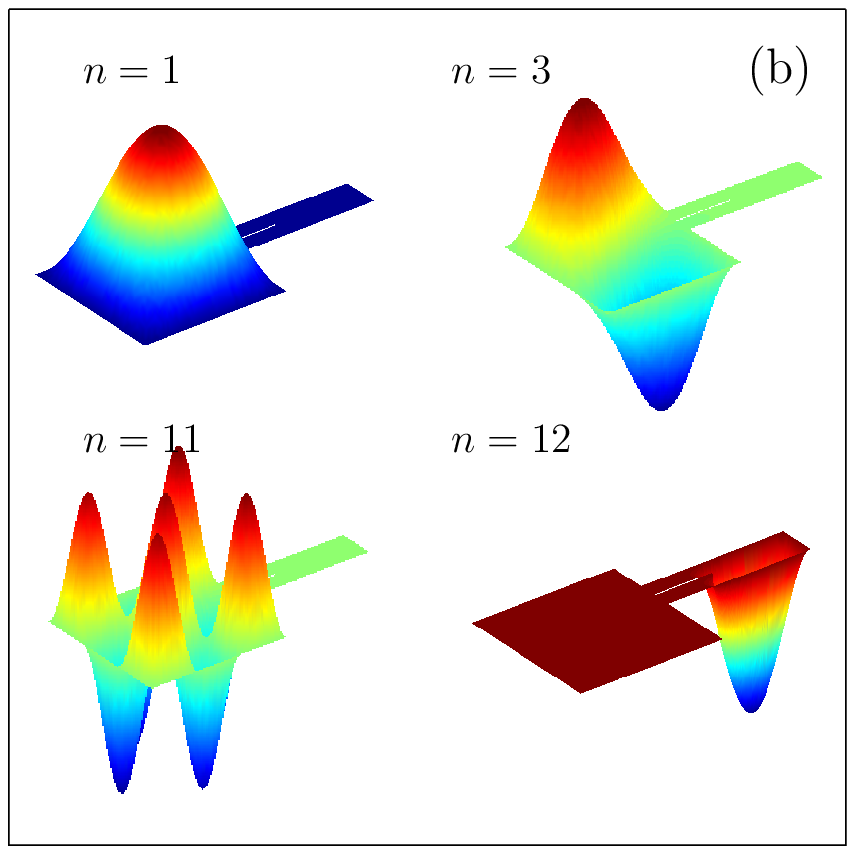}
\end{center}
\caption{
(Color online) The squared $L_2$-norm, $J_n(x_0)$, of three
eigenfunctions with $n = 1, 11, 12$ (symbols) for the branch with a
cut on Fig. \ref{fig:domains}f.  The estimate (\ref{eq:J_ineq3}) with
$\mu = \pi^2/(1/4)^2$ is plotted by solid ($n = 1$) and dash-dotted
($n = 11$) lines.  The hypothetical estimate with $\mu =
\pi^2/(1/8)^2$ for $n = 1$ is shown by dashed line.  The vertical
dotted line indicates the end of the cut.}
\label{fig:J_cut_sym}
\end{figure}

Figure \ref{fig:J_cut_sym}a shows the squared $L_2$-norm, $J_n(x_0)$,
of three eigenfunctions with $n = 1, 11, 12$.  For the first
eigenfunction, we plot two exponential estimates, one with the
rigorous value $\mu = \pi^2/(1/4)^2$ and the other with a hypothetical
value $\mu' = \pi^2/(1/8)^2 = 4\mu$ for a twice narrower branch (if
the cut was complete).  Although the first estimate is of course
applicable for the whole region, it is not sharp.  In turn, the second
estimate is sharp but it works only up to $x_0 < 0.4$.  One can see
that for $x_0 > 0.4$, the slope of $\ln J_1(x_0)$ is given by the
first estimate.  The best estimate would be a combination of these two
but its construction depends on the specific shape of the branch.  It
is worth noting that the transition between two estimates (i.e., the
point $0.4$) does not appear at the end of the cut (here, at $0.5$).
This means that one cannot consider two subregions (with and without
cut) separately in order to develop the individual estimates.

The 11th eigenvalue $\lambda_{11} \approx 177.0$ exceeds $\mu$ so that
the rigorous decay rate is not applicable.  In turn, the use of the
hypothetical value $\mu'$ provides a good estimate up to $x_0 < 0.3$
but fails for $x_0 > 0.3$.  Finally, the 12th eigenfunction has no
exponential decay (in fact, it is localized in the branch).

\subsubsection*{ Tilted and circular branches }

\begin{figure}
\begin{center}
\includegraphics[width=85mm]{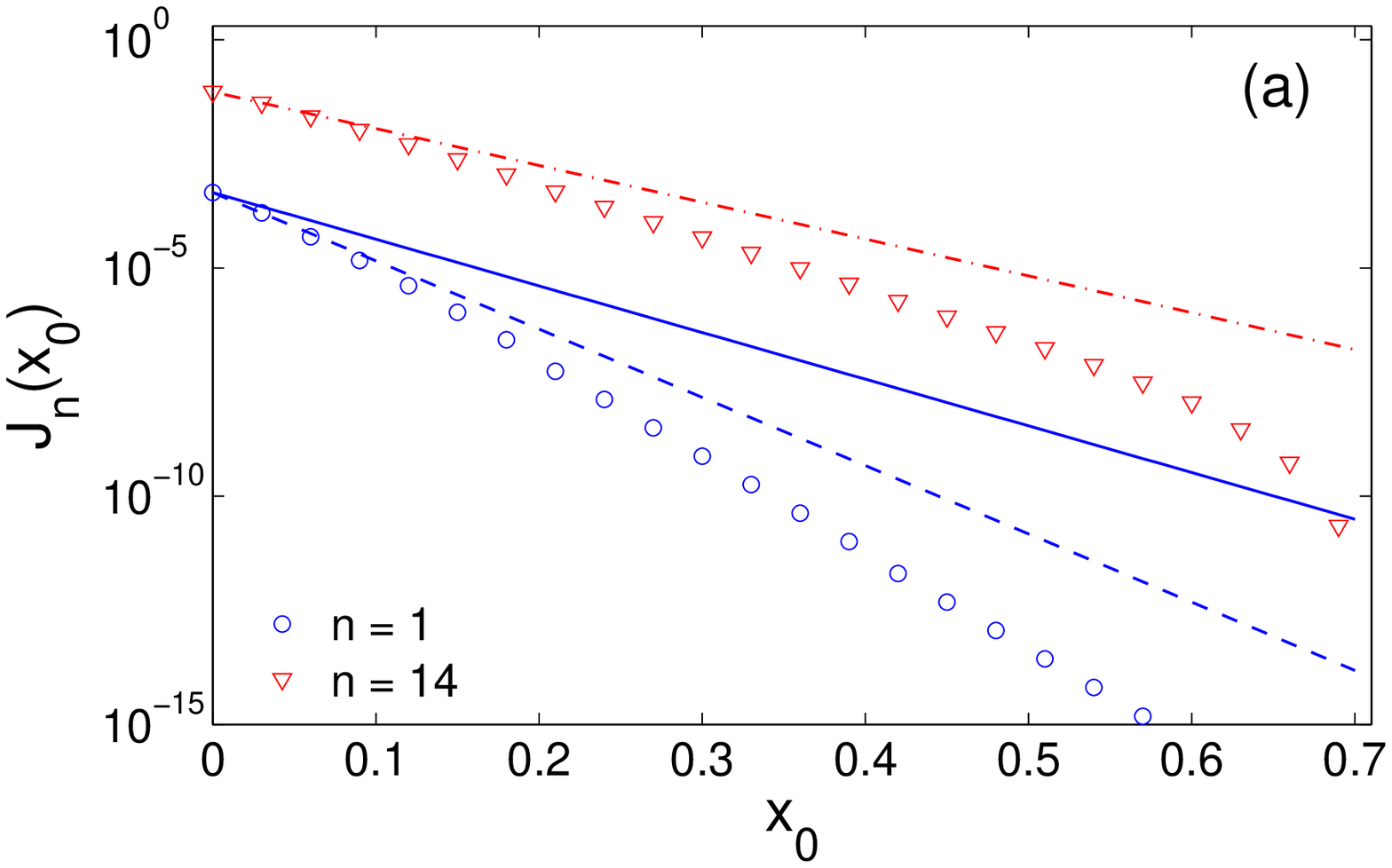}
\includegraphics[width=75mm]{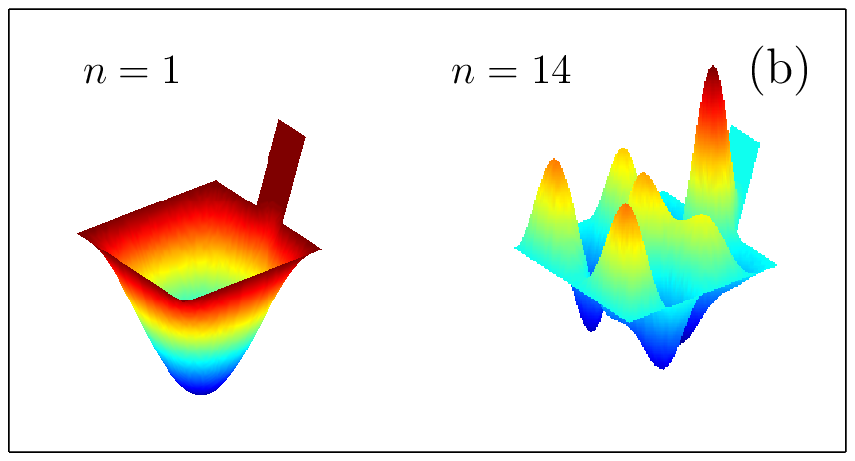}
\end{center}
\caption{
(Color online) The squared $L_2$-norm, $J_n(x_0)$, of two
eigenfunctions with $n = 1, 14$ (symbols) for the tilted branch on
Fig. \ref{fig:domains}g.  The formal estimate (\ref{eq:J_ineq3}) with
$\mu = \pi^2/(1/4)^2$ is plotted by solid line for $n = 1$.  The
improved estimate with $\mu = \pi^2/(1/4/\sqrt{2}))^2$ is shown by
dashed ($n = 1$) and dash-dotted ($n = 14$) lines.  }
\label{fig:J_tilted}
\end{figure}

Another interesting question is the parameterization of the branch
$Q$.  In Sec. \ref{sec:general}, a variable shape of the branch was
implemented through the cross-section profile $\Omega(x)$ where $x$
ranged from $0$ and $a$.  The choice of the $x$ coordinate is
conventional and any other coordinate axis could be used instead of
$x$ (by rotating the domain).  However, the freedom of rotation may
lead to inaccurate estimate for the decay rate.  This point is
illustrated by the domain on Fig. \ref{fig:domains}g with a
rectangular branch tilted by $45^\circ$.  Applying formally the
estimate (\ref{eq:expon}), one expects the exponential decay with $\mu
= \pi^2/(1/4)^2 \approx 157.91$.  At the same time, if the whole
domain was turned clockwise by $45^\circ$ [or, equivalently, if the
branch was parameterized along the axis $x'$ in the direction
$(1,1)$], the decay rate would be set by $\mu =
\pi^2/(1/4/\sqrt{2})^2\approx 315.83$.  It is clear that the behavior
of eigenfunctions does not depend neither on rotations of the domain,
nor on the parameterization of the branch.  Figure \ref{fig:J_tilted}
confirms the exponential decay with the latter threshold $\mu$.

\begin{figure}
\begin{center}
\includegraphics[width=85mm]{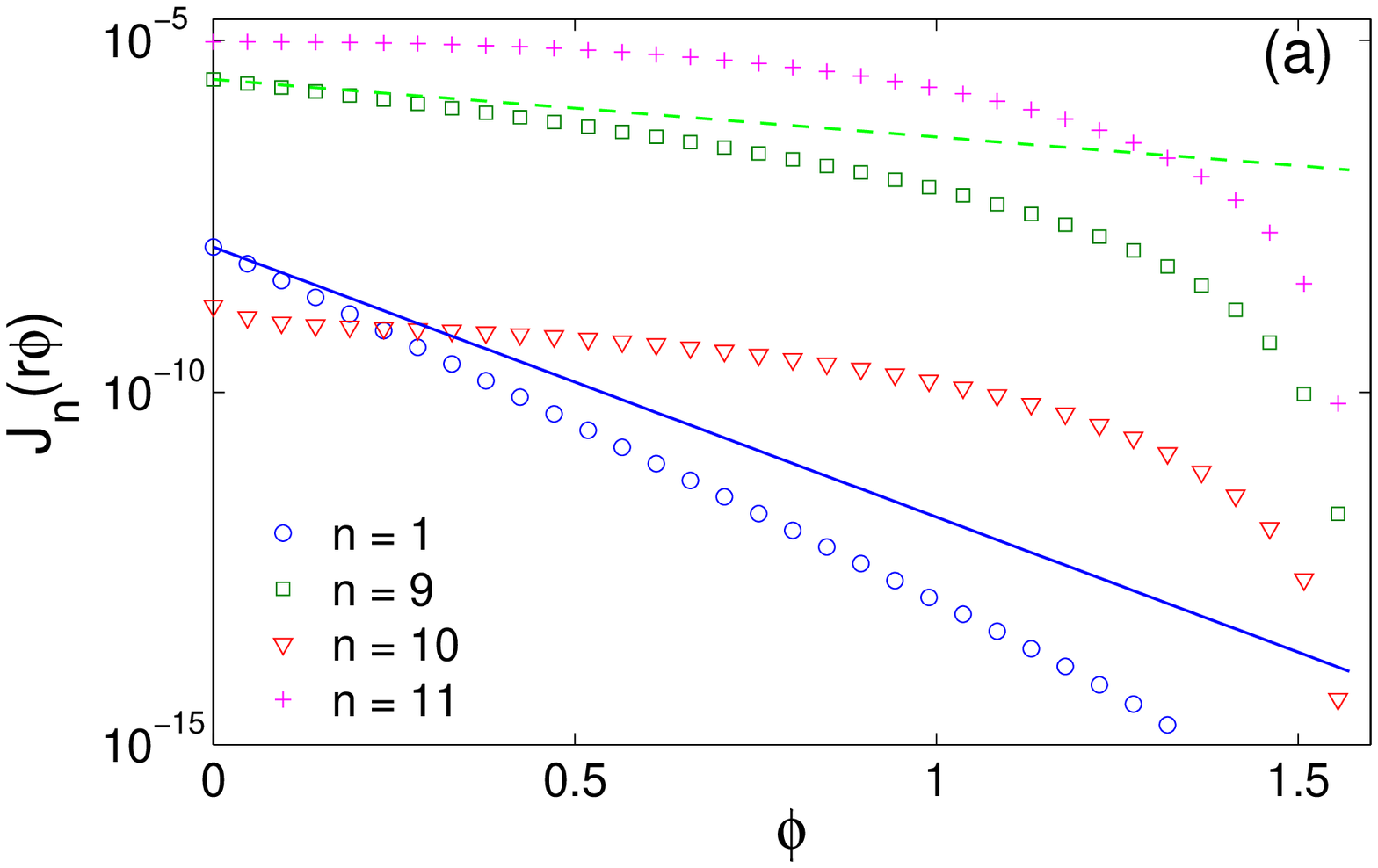}
\includegraphics[width=75mm]{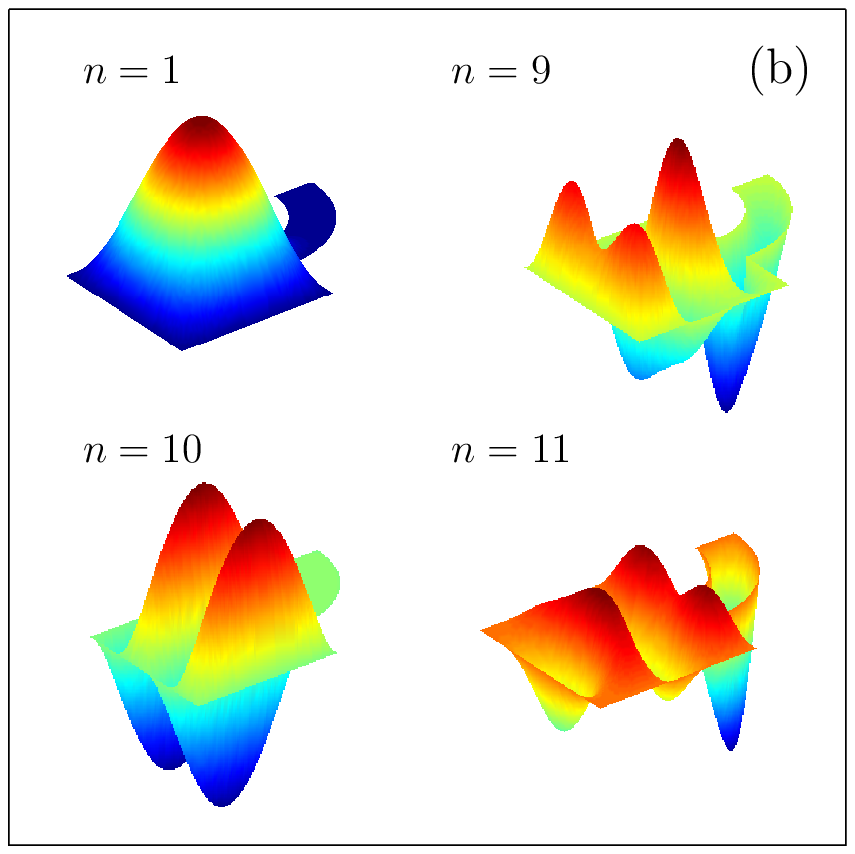}
\end{center}
\caption{
The squared $L_2$-norm, $J_n(x'_0)$, of four eigenfunctions with $n =
1, 9, 10, 11$ (symbols) for the circular branch on
Fig. \ref{fig:domains}h.  The estimate (\ref{eq:J_ineq3}) with $\mu =
\pi^2/(1/4)^2$ is plotted by solid ($n=1$) and dashed ($n=9$)
lines.  The curvilinear coordinate $x_0' = r \phi$ with $r = 3/8$ and
the angle $\phi$ varying from $0$ to $\pi/2$, was used.  }
\label{fig:J_circular}
\end{figure}

A circular branch on Fig. \ref{fig:domains}h is another example, for
which the choice of parameterization is important.  Using the
``conventional'' parameterization by the $x$ coordinate, the largest
cross-section appears at $x = 3/8$, and it is equal to $1/2$ so that
$\mu = \pi^2/(1/2)^2 \approx 39.48$.  Only the first eigenvalue
$\lambda_1$ is below $\mu$ so that the exponential estimate is not
formally applicable to other eigenfunctions.  At the same time, it is
clear that the circular branch of ``true'' width $b$ can be naturally
parameterized by the angle $\phi$ or, equivalently, by an arc, as
illustrated on Fig. \ref{fig:domains}h.  However, there is an
ambiguity in the choice between arcs of various radii (e.g., inner,
outer or middle arcs).  Although the length of all these arcs is
proportional to the angle $\phi$, the proportionality coefficient
enters in the decay rate.  For the results plotted on
Fig. \ref{fig:J_circular}, the inner arc of radius $r = 3/8$ was used.
The squared $L_2$-norm was plotted as a function of the curvilinear
coordinate $x_0' = r\phi$, with $\phi$ varying between $0$ and
$\pi/2$.  For such a curvilinear parameterization, the width of the
branch is constant, $b = 1/4$, so that there are 9 eigenvalues
$\lambda_n$ below $\mu = \pi^2/(1/4)^2$.  For instance, the
exponential decay of the 9th eigenfunction is confirmed on
Fig. \ref{fig:J_circular}.  For circular branches, rigorous estimates
can be derived by writing explicit series representations of
eigenfunctions in a circular annulus and reformulating the analysis
similar to that of Sec. \ref{sec:rectangular} (this analysis is beyond
the scope of the paper).  In general, an extension of the derivation
in Sec. \ref{sec:general} to curvilinear parameterizations is an
interesting perspective.

\subsubsection*{ Branch with a small broadening }

A small broadening in the middle is another interesting perturbation
of the rectangular branch (Fig. \ref{fig:domains}i).  Although this
perturbation is small, the threshold value is reduced to $\mu =
\pi^2/(1/2)^2 \approx 39.48$, instead of the value $\mu' =
\pi^2/(1/4)^2\approx 157.91$ for the rectangular branch.  As a
consequence, the sufficient condition $\lambda_n < \mu$ is satisfied
only for $n = 1$, while for other $n$, the exponential estimate
(\ref{eq:expon}) cannot be applied.  One may thus wonder how do these
eigenfunctions behave in the branch?

\begin{figure}
\begin{center}
\includegraphics[width=85mm]{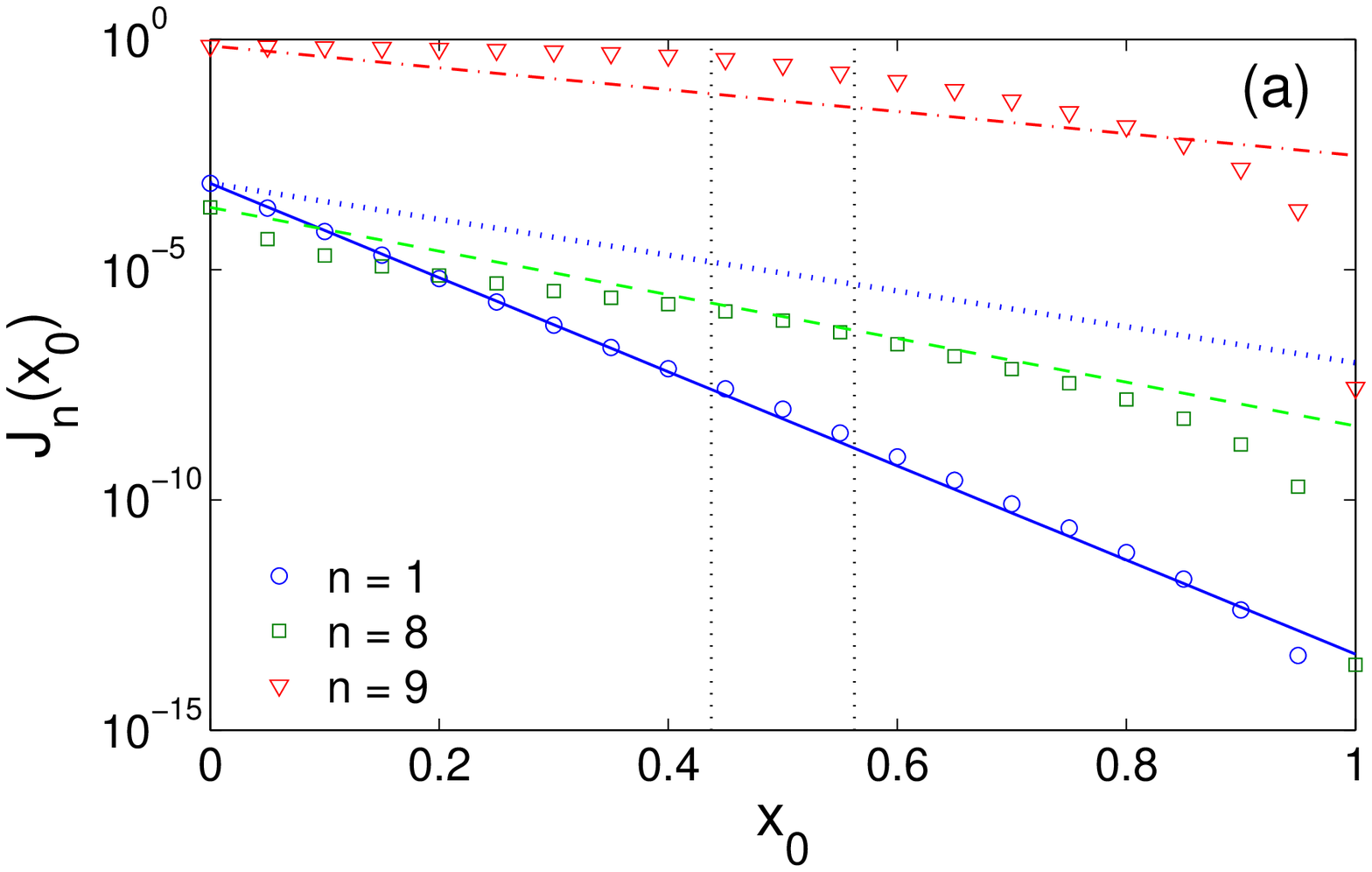}
\includegraphics[width=75mm]{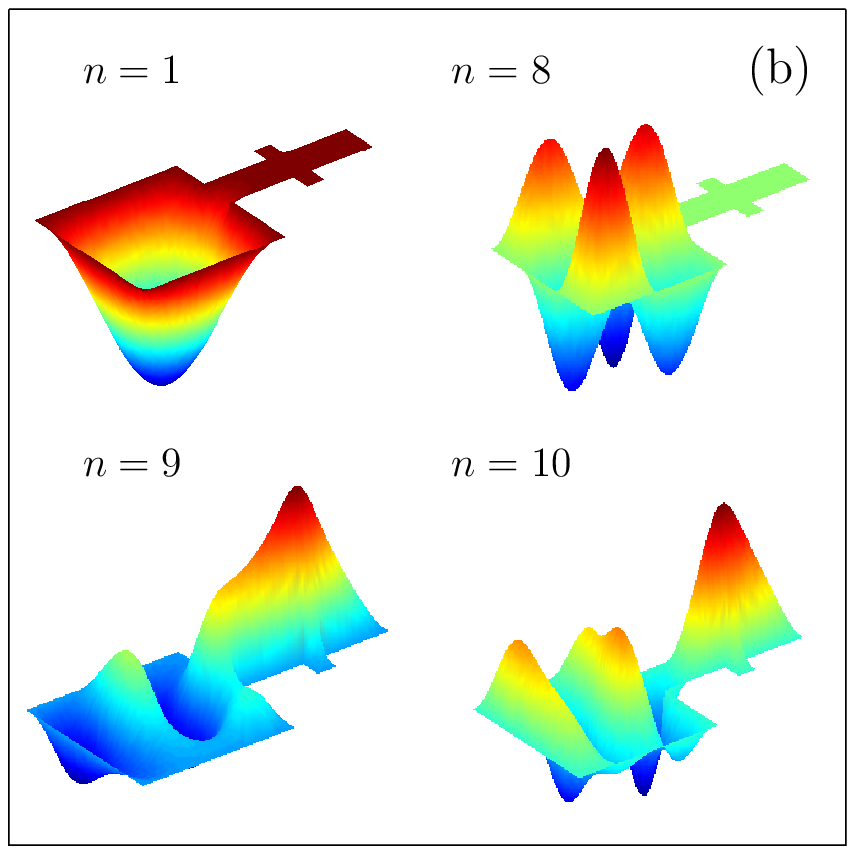}
\end{center}
\caption{
(Color online) The squared $L_2$-norm, $J_n(x_0)$, of three
eigenfunctions with $n = 1, 8, 9$ (symbols) for the branch with a
broadening on Fig. \ref{fig:domains}i.  The estimate
(\ref{eq:J_ineq3}) with $\mu = \pi^2/(1/2)^2$ for $n=1$ is shown by
dotted line.  The hypothetical estimate (\ref{eq:J_ineq3}) with $\mu =
\pi^2/(1/4)^2$ is plotted by solid ($n=1$), dashed ($n=8$) and
dash-dotted ($n = 9$) lines.  Vertical dotted lines indicate the
position of the broadening. }
\label{fig:J_cross_sym}
\end{figure}

We checked that the 8 first eigenfunctions exponentially decay along
the branch, similarly to the rectangular case.  In turn, the 9th, 10th
and some other eigenfunctions do not exhibit this behavior.  Figure
\ref{fig:J_cross_sym} illustrates this result for three eigenfunctions
with $n = 1, 8, 9$.  Since the rigorous estimate is not applicable, we
plot the function $J_n(0) \exp(-2\sqrt{\mu' - \lambda_n} x_0)$ with
$\mu' = \pi^2/(1/4)^2$ as for the rectangular branch.  One can see
that this function correctly captures the behavior of $J_n(x_0)$ but
fails to be its upper bound (there are some regions in which
$J_n(x_0)$ exceeds this function).

In summary, in spite of the fact that the condition $\lambda < \mu$ is
not satisfied, the presence of a small broadening does not
significantly affect the exponential decay of the first eigenfunctions
but the upper bound is not valid.

\subsubsection*{ Bifurcating branch }

After describing the eigenfunctions in a single branch, one may wonder
about their properties in a general pore network which consists of
basic domains (``pores'') connected by branches.  In spite of numerous
potential applications for diffusive transport in porous media, very
little is known about this challenging problem.  A dumbbell domain
with a thin ``handle'' is probably the most studied case
\cite{Beale73,Gadylshin93,Gadylshin94,Gadylshin05}.  In particular, 
the method of matching asymptotic series yields the asymptotic
estimates when the diameter of the channel is considered as a small
parameter.  

\begin{figure}
\begin{center}
\includegraphics[width=85mm]{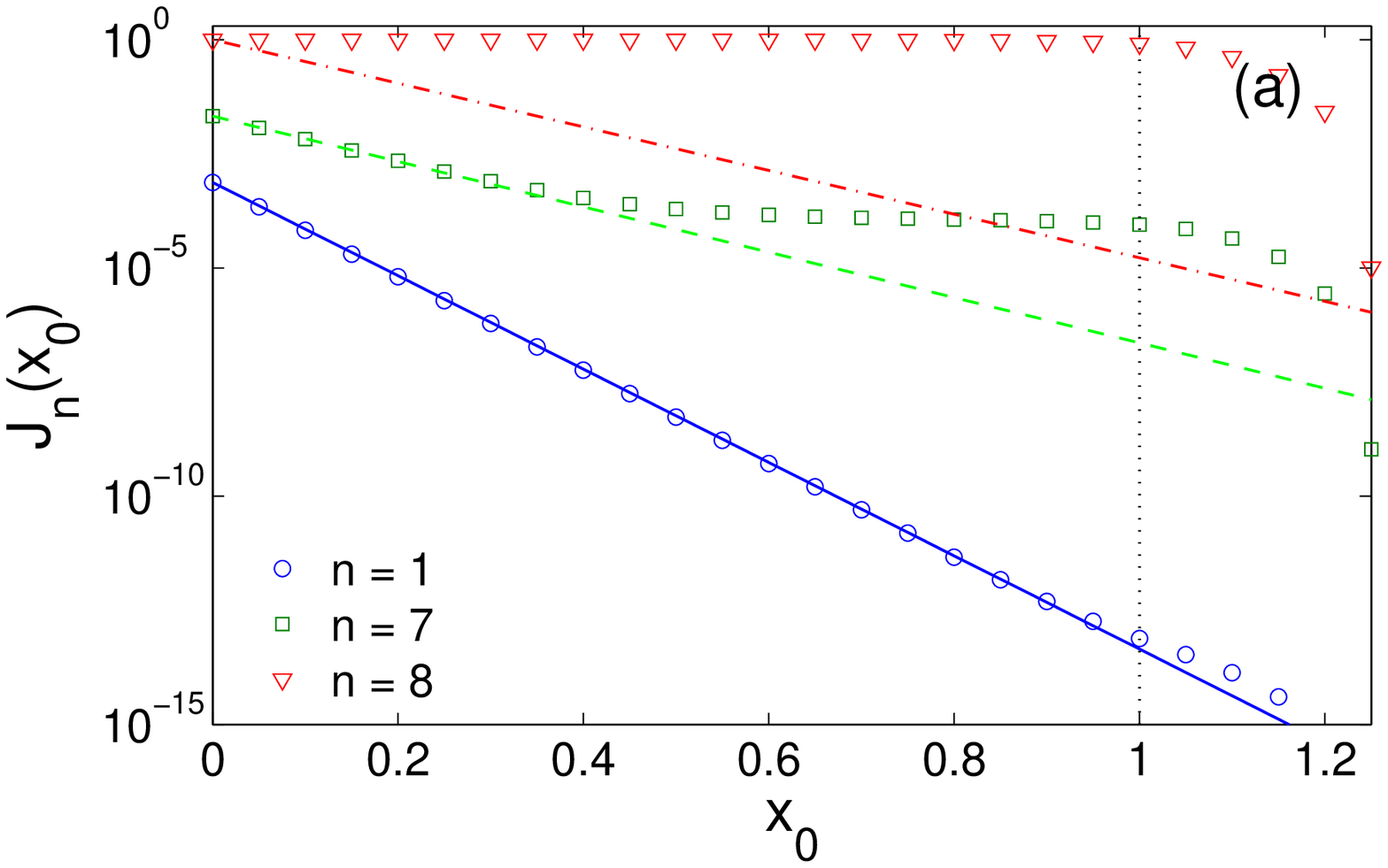}
\includegraphics[width=75mm]{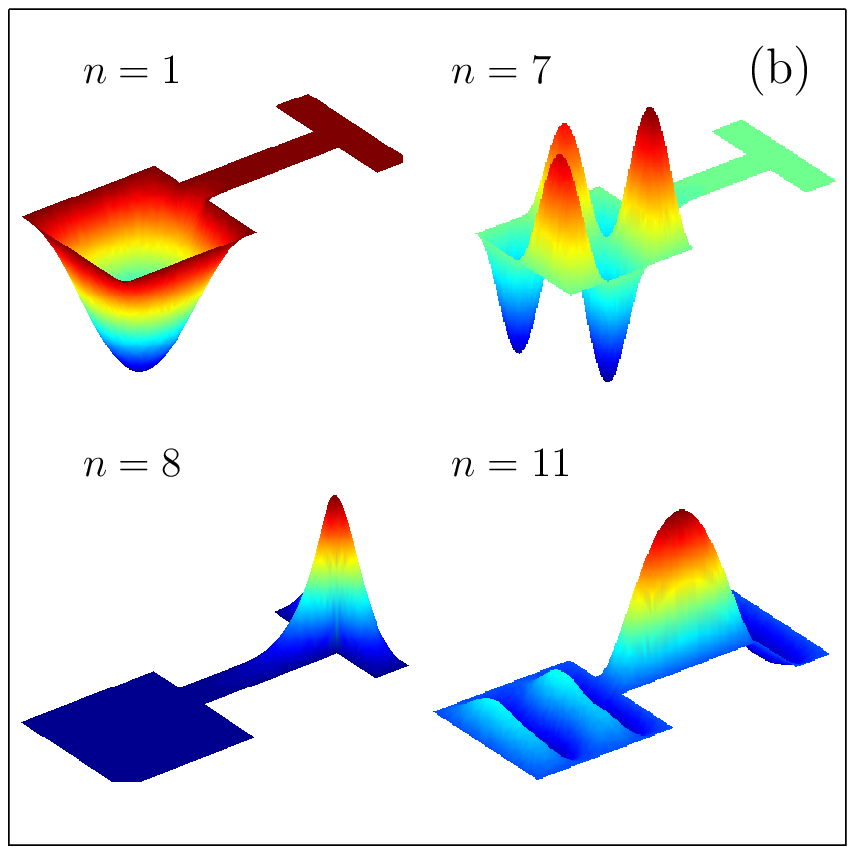}
\end{center}
\caption{
(Color online) The squared $L_2$-norm, $J_n(x_0)$, of three
eigenfunctions with $n = 1, 7, 8$ (symbols) for the bifurcating branch
on Fig. \ref{fig:domains}j.  The hypothetical estimate
(\ref{eq:J_ineq3}) with $\mu = \pi^2/(1/4)^2$ is plotted by solid ($n
= 1$), dashed ($n = 7$) and dash-dotted ($n = 8$) lines.  The vertical
dotted line indicates the position of bifurcation. }
\label{fig:J_bifurcate_sym}
\end{figure}

In order to illustrate the related difficulties, we consider a
rectangular branch which bifurcates into two rectangular branches
(Fig. \ref{fig:domains}j).  For this example, the largest
cross-section length is $1$ so that $\mu = \pi^2\approx 9.87$ and the
exponential estimate is not applicable.  Figure
\ref{fig:J_bifurcate_sym} shows $J_n(x_0)$ for three eigenfunctions
with $n = 1, 7, 8$ and the hypothetical estimate with $\mu' =
\pi^2/(1/4)^2$ as for the rectangular branch.  The first eigenfunction
is well estimated by the exponential function
$J_1(0)\exp(-2\sqrt{\mu'-\lambda_1}x_0)$ for practically the whole
length of the branch ($x_0 < 1$), with a small deviation at the
bifurcation region ($x_0 > 1$).  Similar behavior is observed for
other eigenfunctions up to $n = 7$.  The larger the index $n$, the
earlier the deviation from the exponential estimate appears (e.g., for
$n = 7$, the estimate works only for $x_0 < 0.3$).  In turn, the 8th
eigenfunction has no exponential decay in the branch as it is
localized in the bifurcation region.  At the same time, the
corresponding eigenvalues are close: $\lambda_7 \approx 125.20$ and
$\lambda_8 \approx 127.67$.  This illustrates how significant may be
the difference in properties of two consecutive eigenfunctions.  We
also emphasize on the difficulty of distinguishing these cases in
general.

Surprisingly, an exponential decay can be recovered for thinner
bifurcations of arbitrary length.  Although this example may look
specific, the arguments provided below are of general interest.  In
order to prove the exponential decay of the squared norm $J(x_0)$ in
Eq. (\ref{eq:J_num}), we will establish the inequalities similar to
those for $I(x_0)$ of Sec. \ref{sec:general}.  Replacing the norm
$I(x_0)$ by the weaker norm $J(x_0)$ allows one to produce more
accurate estimates.  Although we bear in mind the example shown on
Fig. \ref{fig:domains}j, the arguments are rather general and
applicable in higher dimensions.  By definition, one has
\begin{equation*}
J'(x_0) = - I(x_0) = - \int\limits_{\Omega(x_0)} u^2(x_0,\y) d\y < 0 
\end{equation*}
and
\begin{equation*}
\begin{split}
J''(x_0) & = - 2\int\limits_{\Omega(x_0)} u \frac{\partial}{\partial x} u d\y  
 = 2\int\limits_{Q(x_0)} \bigl[(\nabla u, \nabla u) + u \Delta u\bigr] dx d\y ,  \\
\end{split}
\end{equation*}
where the Green's formula, the identity $\partial/\partial x = -
\partial/\partial n$ on $\Omega(x_0)$ and Dirichlet boundary condition
on $\partial Q(x_0)\backslash \Omega(x_0)$ were used.  The first
integral can be estimated by splitting $Q(x_0)$ into two subdomains:
$Q_1(x_0)$ (a part of the rectangular branch) and $Q_2$ (the
bifurcation):
\begin{equation*}
\int\limits_{Q(x_0)} (\nabla u, \nabla u) dx d\y  \geq \mu \int\limits_{Q_1(x_0)} u^2 dx d\y  +  \int\limits_{Q_2} (\nabla u, \nabla u) dx d\y ,
\end{equation*}
where the Friedrichs-Poincar\'e inequality was used for $Q_1(x_0)$,
with the smallest $\mu$ over $Q_1(x_0)$ (for the rectangular branch on
Fig. \ref{fig:domains}j, $\mu = \pi^2/b^2$).  The separate analysis of
two subdomains allowed us to exclude large cross-sections of the
bifurcation $Q_2$ from the computation of $\mu$.

In order to estimate the second integral over $Q_2$, we note that
\begin{equation*}
\nu_1 = \inf\limits_{v\in H^1(Q_2),~ v|_\Gamma = 0,~ v\ne 0} \frac{(\nabla v, \nabla v)}{(v,v)}
\end{equation*}
is the smallest eigenvalue of the Laplace operator in the rectangle
$Q_2$ with Dirichlet boundary condition on the top, bottom and right
segments (denoted by $\Gamma$) and Neumann boundary condition on the
left segment.  Since the eigenfunction $u$ also belongs to $H^1(Q_2)$,
one has
\begin{equation*}
(\nabla u, \nabla u)_{L_2(Q_2)} \geq \nu_1 (u,u)_{L_2(Q_2)} ,
\end{equation*}
from which
\begin{equation*}
J''(x_0) \geq 2(\mu-\lambda)\int\limits_{Q_1(x_0)} u^2 dx d\y + 2(\nu_1-\lambda)\int\limits_{Q_2} u^2 dx d\y   .
\end{equation*}
If $\nu_1 > \mu$, we finally obtain
\begin{equation*}
J''(x_0) \geq 2(\mu-\lambda)\int\limits_{Q(x_0)} u^2 dx d\y = 2(\mu-\lambda) J(x_0).
\end{equation*}
As shown in Sec. \ref{sec:general}, this inequality implies an
exponential decay of $J(x_0)$.

If $Q_2$ is a rectangle of height $h$ and width $w$ with Dirichlet
boundary condition on the top, bottom and right segments and Neumann
boundary condition on the left segment, the smallest eigenvalue is
$\nu_1 = \pi^2/h^2 + \pi^2/(2w)^2$.  If the branch $Q_1$ is also a
rectangle of width $b$ (as shown on Fig. \ref{fig:domains}j), one has
$\mu = \pi^2/b^2$.  The condition $\nu_1 > \mu$ reads as $(b/h)^2 +
(b/(2w))^2>1$.  For instance, this condition is satisfied for any $h$
if the bifurcation width $w$ is smaller than $b/2$.  In particular,
this explains that a small broadening of width $b/2$ shown on
Fig. \ref{fig:domains}i does not degrade the exponential decay of
$J_n(x_0)$ for $x_0 < 0.4375$.  In turn, the bifurcation shown on
Fig. \ref{fig:domains}j with $h = 1$ and $w = b = 1/4$ does not
fulfill the above condition as $(b/h)^2 + (b/(2w))^2 = 1/16 + 1/4 <
1$.

\subsection{ Infinite branches }
\label{sec:infinite}

In the derivation of Sec. \ref{sec:general}, there was no condition on
the length $a$ of the branch.  Even for an infinite branch, the
exponential decay is valid once the condition (\ref{eq:assumpt}) is
satisfied.  However, when the branch is infinite, the Laplace operator
eigenspectrum is not necessarily discrete so that $L_2$-normalized
eigenfunctions may not exist.  A simple counter-example is a
semi-infinite strip $D = [0,\infty)\times [0,\pi]$ for which functions
$\sinh(\sqrt{n^2 - \lambda} x) \sin(ny)$ satisfy the eigenvalue
problem (\ref{eq:eigenproblem}) but their $L_2(D)$ norms are infinite.

As shown by Rellich, if the first eigenvalue $\mu_1(x)$ in the
cross-section $\Omega(x)$ of an infinite branch $Q$ goes to infinity
as $x\to\infty$, then there exist infinitely many $L_2$-normalized
eigenfunctions \cite{Rellich}.  In two dimensions, $\mu_1(x) =
\pi^2/\ell(x)^2$ is related to the length $\ell(x)$ of the largest
interval in the cross-section $\Omega(x)$.  The condition
$\mu_1(x)\to\infty$ requires thus $\ell(x)\to 0$.  At the same time,
the number of intervals in $\Omega(x)$ can increase with $x$ so that
the ``total width'' of the branch may arbitrarily increase.

For infinite decreasing branches, eigenfunctions can be shown to decay
{\it faster} than an exponential with any decay rate.  In fact,
although the threshold $\mu$ was defined in (\ref{eq:mu}) as the
smallest $\mu_1(x)$ over all cross-sections of the branch, the
separation into a basic domain and a branch was somewhat arbitrary.
If the separation occurs at $x_0$ (instead of $0$), one gets
\begin{equation*}
I(x) \leq I(x_0) \exp[-2\sqrt{\mu(x_0) - \lambda}~ (x - x_0)]  \quad x \geq x_0 ,
\end{equation*}
where the new threshold $\mu(x_0) = \inf\limits_{x_0 < x} \mu_1(x)$
increases with $x_0$, while the prefactor $I(x_0)$ also decays
exponentially with $x_0$ according to Eq. (\ref{eq:expon}).  Since the
above estimate is applicable for any $x_0$, one can take $x_0$ to be a
slowly increasing function of $x$ [its choice depends on $\mu(x_0)$]
that would result in a faster-than-exponential decay of $I(x)$.  This
result is a consequence of the condition $\mu_1(x)\to\infty$ as $x$
goes to infinity.

\subsection{ Three-dimensional domains }

The exponential estimate (\ref{eq:expon}) becomes still more
interesting in three (and higher) dimensions.  While any cross-section
$\Omega(x)$ was a union of intervals in two dimensions, the shape of
cross-sections in three dimensions can vary significantly (e.g., see
Fig. \ref{fig:domain}c).  Whatever the shape of the branch is, the
only relevant information for the exponential decay is the smallest
eigenvalue $\mu$ in all cross-sections $\Omega(x)$.  For instance, for
a rectangular profile of the branch, $\Omega(x) = [0,b(x)]\times
[0,c(x)]$, the first eigenvalue $\mu_1(x) = \frac{\pi^2}{b(x)^2} +
\frac{\pi^2}{c(x)^2}$ can remain bounded from below by some $\mu$ even
if one of the sides $b(x)$ or $c(x)$ grows to infinity.  This means
that eigenfunctions may exponentially decay even in infinitely growing
branches.

\subsection{ Neumann boundary condition }

\begin{figure}
\begin{center}
\includegraphics[width=75mm]{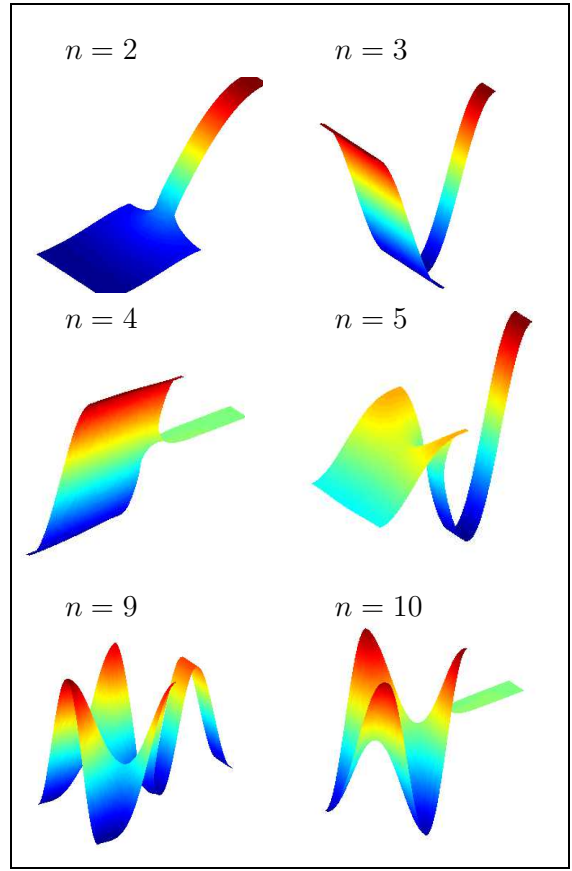}
\end{center}
\caption{
Six eigenfunctions with $n = 2,3,4,5,9,10$ for the rectangular branch
(Fig. \ref{fig:domains}a) with Neumann boundary condition (the
fundamental eigenfunction with $n=1$ is constant and not shown).  }
\label{fig:eigen_N}
\end{figure}

The theoretical derivation in Sec. \ref{sec:general} essentially
relies on the Dirichlet boundary condition on the branch boundary:
$u|_{\dc Q} =0$.  This condition can be interpreted, e.g., as a rigid
fixation of a vibrating membrane at the boundary, or as a perfect
absorption of diffusing particles at the boundary.  The opposite case
of free vibrations of the membrane or a perfect reflection of the
particles is described by Neumann boundary condition, $\partial
u/\partial n|_{\dc Q} =0$.  Although the eigenvalue problem may look
similar, the behavior of eigenfunctions is different.  In particular,
an extension of the results of Sec. \ref{sec:general} fails even in
the simplest case of a rectangular branch, as illustrated on
Fig. \ref{fig:eigen_N}.  Although the 8 first eigenvalues $\lambda_n$
are below $\mu = \pi^2/(1/4)^2$, only some of them decay exponentially
(e.g., with $n = 4$).  This decay seems to be related to the
reflection symmetry of the domain.  Since the eigenfunctions with $n =
4,10$ are anti-symmetric, they are $0$ along the horizontal symmetry
line.  One can therefore split the domain into two symmetric
subdomains and impose the Dirichlet boundary condition on the
splitting line.  Although an exponential decay may be expected for
this new problem, its mathematical justification is beyond the scope
of the paper.

On the other hand, the derivation of the exponential estimate in
Sec. \ref{sec:general} does not use the boundary condition imposed on
the basic domain $V$.  As a consequence, the estimate is applicable
for arbitrary boundary condition on $V$ which guarantees the Laplace
operator in the whole domain $D$ to be self-adjoint.

\subsection{ Expelling from the branch }

It is important to stress that the ``smallness'' of an eigenfunction
in the branch and its exponential decay are different notions which
should not to be confused.  In fact, the eigenfunction can be small in
the branch either due to a rapid exponential decay, or because of the
small constant $I_n(0)$ or $J_n(0)$ in front of the estimate.  For
instance, Fig. \ref{fig:J_circular}a shows the behavior of
$J_{10}(x_0)$ without an exponential decay, but the eigenfunction is
nevertheless small (Fig. \ref{fig:J_circular}b).  Another example on
Fig. \ref{fig:rectangle_J}a with $n = 9$ illustrates the opposite
situation: the eigenfunction decays exponentially along the branch but
it does not look small.

\section*{ Conclusion }

We have studied the behavior of the Laplace operator eigenfunctions in
a large class of domains composed of a basic domain of arbitrary shape
and a branch $Q$ which can be parameterized by a variable profile
$\Omega(x)$.  We have rigorously proved that each eigenfunction whose
eigenvalue $\lambda$ is smaller than the threshold $\mu =
\inf\{\mu_1(x)\}$, exponentially decays inside the branch, where
$\mu_1(x)$ is the first eigenvalue of the Laplace operator in the
cross-section $\Omega(x)$.  In general, the decay rate was shown to be
at least $\sqrt{2}\sqrt{\mu-\lambda}$.  For non-increasing branches,
the decay rate $2\sqrt{\mu - \lambda}$ was derived and shown to be
sharp for an appropriate parameterization of the branch.  The
exponential estimate is applicable in any dimension and for finite and
infinite branches.  In the latter case, the condition
$\mu_1(x)\to\infty$ as $x\to\infty$ is imposed to ensure the existence
of $L_2$-normalized eigenfunctions.  Since the derivation did not
involve any information about the basic domain $V$, the exponential
estimate is applicable for arbitrary $V$ with any boundary condition
on $\dc V$ for which the Laplace operator in $D$ is still
self-adjoint.  In turn, the Dirichlet boundary condition on the branch
boundary was essential.  Note that the mathematical methods of the
paper can be adapted for studying eigenfunctions for various spectral
problems or other types of domains.

The numerical simulations have been used to illustrate and extend the
theoretical results.  It was shown that the sufficient condition
$\lambda < \mu$ is not necessary, i.e., the eigenfunctions may
exponentially decay even if $\lambda > \mu$.  However, in this case,
the decay rate and the range of its applicability strongly depend on
the specific shape of the branch.  For all numerical examples, the
sharp decay rate $2\sqrt{\mu-\lambda}$ was correct, even if the
condition (\ref{eq:cond}) for non-increasing branches was not
satisfied.  In future, it is tempting either to relax this condition,
or to find counter-examples, for which the sharp decay is not
applicable.

\section*{Acknowledgment}

This work has been partly supported by the RFBR N 09-01-00408a grant
and the ANR grant ``SAMOVAR''.

\begin{appendix}

\section{ Estimate for rectangular branch }
\label{sec:rectangle}

From the inequality $\sinh x \leq \cosh x$, Eq. (\ref{eq:energy}) is
bounded as
\begin{equation*}
||\nabla u||^2_{L_2(Q(x_0))} \leq \frac{b}{2} \sum \limits_{n=1}^{\infty} c_n^2  
\biggl[(\frac{\pi}{b} n)^2 + \gamma_n^2\biggr] \int\limits_{x_0}^{a} \cosh^2 (\gamma_n (a-x)) dx .
\end{equation*}
The last integral is estimated as
\begin{equation*}
\begin{split}
& \int\limits_{x_0}^{a} \cosh^2 (\gamma_n (a-x)) dx \leq \int\limits_{x_0}^{a} e^{2\gamma_n (a-x)} dx = \\
& \frac{e^{2\gamma_n a}}{2\gamma_n} \bigl(e^{-2\gamma_n x_0} - e^{-2\gamma_n a}\bigr) \leq 
\frac{e^{2\gamma_n a}}{2\gamma_n}  e^{-2\gamma_n x_0} \leq \frac{e^{2\gamma_n a}}{2\gamma_n}  e^{-2\gamma_1 x_0} , \\
\end{split}
\end{equation*}
where we used the inequality $\cosh x \leq e^x$ for $x\geq 0$ and the
fact that $\gamma_n = \sqrt{\pi^2 n^2/b^2 - \lambda}$ increases with
$n$.  We have then
\begin{equation*}
||\nabla u||^2_{L_2(Q(x_0))} \leq \frac{b}{2} e^{-2\gamma_1 x_0} 
\sum \limits_{n=1}^{\infty} c_n^2 \biggl[\frac{\pi^2}{b^2} \frac{n^2}{2\gamma_n} + \frac{\gamma_n}{2}\biggr] e^{2\gamma_n a}  .
\end{equation*}
Writing two inequalities: 
\begin{equation*}
\begin{split}
\gamma_n & = \sqrt{\pi^2 n^2/b^2 - \lambda} \leq \frac{\pi}{b} n ,  \\
\frac{n^2}{\gamma_n} & = \frac{n}{(\pi/b)\sqrt{1-\lambda b^2/(\pi^2 n^2)}} \leq C_1 n , \\
\end{split}
\end{equation*}
where $C_1$ is a constant, one gets an upper bound in the order of $n$
for the expression in large brackets.  Finally, we have an estimate
for $e^{2\gamma_n a}$ as
\begin{equation*}
\sinh^2(2\gamma_n a) = \frac{e^{2\gamma_n a}}{4} (1 - e^{-2\gamma_n a})^2 \geq \frac{e^{2\gamma_n a}}{4} (1 - e^{-2\gamma_1 a})^2 ,
\end{equation*}
from which
\begin{equation*}
e^{2\gamma_n a} \leq C_2 \sinh^2(2\gamma_n a) ,
\end{equation*}
with a constant $C_2 = 4/(1-e^{-2\gamma_1 a})^2$.  Bringing together
these inequalities, we get the estimate (\ref{eq:aux10}).

\subsection*{ Trace theorem }

The trace theorem implies \cite{Lions} that the series 
\begin{equation*}
f(x) \equiv \sum\limits_{n=1}^\infty n \bigl(u(x,y), \sin (\pi n y/b)\bigr)^2_{L_2(0, b)} , 
\end{equation*}
which is equivalent to the squared norm of $u(x,y)$ in the Sobolev
space $H_{(0,b)}^{1/2}$, may be estimated from above by the norm
$||\nabla u||^2_{L_2(D)}$.  For completeness, we provide the proof for
our special case.

For a fixed $x$, we denote 
\begin{equation*}
X_n(x) \equiv \bigl(u(x,y), \sin (\pi n y/b)\bigr)_{L_2(0,b)} 
\end{equation*}
the Fourier coefficients of the function $u(x,y)$:
\begin{equation*}
u(x,y) = \frac{2}{b}\sum\limits_{n=1}^\infty X_n(x) \sin (\pi n y/b) .
\end{equation*}
On one hand, starting from $X_n(a) = 0$, one gets
\begin{equation*}
X^2_n(x) = \left| \int \limits_x^a (X^2_n)' dx_1 \right| = 2 \left| \int \limits_x^a X_n X'_n dx_1 \right|  ,
\end{equation*}
while the Cauchy inequality implies
\begin{equation*}
2 \left| \int \limits_x^a X_n X'_n dx_1 \right| \leq 2 ||X_n||_{L_2(0,a)}||X'_n||_{L_2(0,a)} .
\end{equation*}
The inequality $2\alpha \beta \leq \alpha^2 + \beta^2$ yields
\begin{equation*}
\begin{split}
2(\pi n/b) & ||X_n||_{L_2(0,a)}||X'_n||_{L_2(0,a)} 
 \leq ||X'_n||^2_{L_2(0,a)} + (\pi n/b)^2 ||X_n||^2_{L_2(0,a)} , \\
\end{split}
\end{equation*}
from which
\begin{equation*}
(\pi n/b) X^2_n(x) \leq ||X'_n||^2_{L_2(0,a)} + (\pi n/b)^2 ||X_n||^2_{L_2(0,a)} .
\end{equation*}
On the other hand, we write explicitly the energetic norm of $u$:
\begin{equation*}
||\nabla u||^2_{L_2(Q)} = \frac{2}{b} \sum\limits_{n=1}^{\infty} \biggl(||X'_n||^2_{L_2(0,a)} + (\pi n/b)^2 
||X_n||^2_{L_2(0,a)}\biggr),
\end{equation*}
from which
\begin{equation*}
f(x) = \sum\limits_{n=1}^\infty n X^2_n(x) \leq \frac{b^2}{2\pi} ||\nabla u||^2_{L_2(Q)} \leq \frac{b^2}{2\pi} ||\nabla u||^2_{L_2(D)} .
\end{equation*}

Since the coefficients $X_n(x)$ and $c_n$ are related as
\begin{equation*}
X_n(x) = \frac{b}{2}~ c_n \sinh(\gamma_n(a-x)) ,
\end{equation*}
the substitution of $x = 0$ into this equation yields
\begin{equation}
\begin{split}
\sum\limits_{n=1}^\infty n c_n^2 \sinh^2(\gamma_n a) & = \frac{4}{b^2} \sum\limits_{n=1}^\infty n X_n(0)^2 = \frac{4}{b^2} f(0) 
 \leq \frac{2\lambda}{\pi}  ||u||^2_{L_2(D)} = \frac{2\lambda}{\pi} ||u||^2_{L_2(D)} . \\
\end{split}
\end{equation}

\section{ Several classical results }

For completeness, we recall the derivation of several classical
results \cite{Lions,Glazman} which are well known for spectral
analysts but may be unfamiliar for other readers.

\subsection{ Rayleigh's principle }
\label{sec:rayleigh}
\label{sec:Friedrichs}

Let us start with the first eigenvalue $\lambda_1$ of the problem
(\ref{1}) which can be found as
\begin{equation}
\label{eq:lambda1}
\lambda_1 = \inf \limits_{v \in \overset{\circ}{H^1}} \frac{(\nabla v, \nabla v)_{L_2(D)}}{(v,v)_{L_2(D)}}  ,
\end{equation}
where $\overset{\circ}{H^1} = \{v \in L_2(D),~ \partial v/\partial x_i
\in L_2(D),~ i=1,...,n+1,~ v|_{\dc D} = 0 \}$.  Denoting $\phi_1$ the first
eigenfunction in Eq. (\ref{2}), one takes
\begin{equation*}
v = \begin{cases} \phi_1, \quad (x,\y) \in V, \cr 0 , \hskip 5.5mm (x,\y) \notin V, \end{cases}
\end{equation*}
as a trial function in Eq. (\ref{eq:lambda1}) to obtain
\begin{equation*}
\lambda_1 < \frac{(\nabla \phi_1, \nabla \phi_1)_{L_2(V)}}{(\phi_1, \phi_1)_{L_2(V)}} = \kappa_1 ,
\end{equation*}
i.e., the first eigenvalue $\lambda_1$ in the whole domain $D$ is
always smaller than the first eigenvalue $\kappa_1$ in its subdomain
$V$.  More generally, if there are $n$ eigenvalues $ \kappa_1 \leq
\dots \leq \kappa_n \leq \mu$ then there exist $n$ eigenvalues
$\lambda_1 \leq \dots \leq \lambda_n < \mu$.

Note that the Friedrichs-Poincar\'e inequality (\ref{eq:Poincare})
follows from (\ref{eq:lambda1}).

\subsection{ Rellich's identity }
\label{sec:Rellich}

Let $u$ be an eigenfunction which satisfies the equation
\begin{equation*}
\Delta u + \lambda u = 0 \quad (x, \y) \in D, \quad u|_{\dc D} = 0 .
\end{equation*}
We multiply this equation by $\frac{\dc u}{\dc x}$ and integrate over
the domain $Q(x_0)$ defined by Eq. (\ref{eq:Qx0}):
\begin{equation}
\label{eq:aux4}
\int \limits_{Q(x_0)} \frac{\dc u}{dx} \Delta u ~ dx d\y + \lambda \int \limits_{Q(x_0)} u \frac{\dc u}{dx} ~dx d\y = 0 .
\end{equation}
The second integral can be transformed as
\begin{equation}
\label{eq:aux6}
\begin{split}
& \lambda \int \limits_{Q(x_0)} u \frac{\dc u}{dx} ~dx d\y = 
\frac{\lambda}{2} \int\limits_{Q(x_0)} \biggl(\frac{\dc}{\dc x} u^2\biggr) dx d\y \\
& = - \frac{\lambda}{2}  \int\limits_{\dc Q(x_0)} u^2(x_0, \y) (\e_x, \n) dS =
- \frac{\lambda}{2}  \int\limits_{\Omega(x_0)} u^2(x_0, \y) d\y , \\
\end{split}
\end{equation}
where $\n = \n(S)$ is the unit normal vector at $S\in \dc Q(x_0)$ and
the boundary condition $u_{\dc D} = 0$ was used on $\Gamma(x_0) = \dc
Q(x_0)\backslash \Omega(x_0)$.

Using the Green's formula, the first integral in Eq. (\ref{eq:aux4})
can be transformed to
\begin{equation}
\label{eq:aux5}
\int\limits_{Q(x_0)} \hspace*{-2mm} \frac{\dc u}{\dc x} \Delta u dx d\y 
= \hspace*{-2mm} \int\limits_{\dc Q(x_0)} \hspace*{-2mm} \frac{\dc u}{\dc x} \frac{\dc u}{\dc n} dS  
- \int\limits_{Q(x_0)} \biggl(\nabla u, \nabla \frac{\dc u}{\dc x}\biggr) dx d\y .
\end{equation}
The first integral over $\dc Q(x_0)$ can be split in two terms:
\begin{equation*}
\int\limits_{\dc Q(x_0)} \frac{\dc u}{\dc x} \frac{\dc u}{\dc n} dS = 
\int\limits_{\Gamma(x_0)} \frac{\dc u}{\dc x} \frac{\dc u}{\dc n} dS - \int\limits_{\Omega(x_0)} \biggl(\frac{\dc u}{\dc x}\biggr)^2 d\y,
\end{equation*}
where $\partial/\partial n = (\n, \nabla)$ is the normal derivative
pointing outwards the domain, and the sign minus appears because
$\partial u/\partial n = - \partial u/\partial x$ at $\Omega(x_0)$.

The second integral in Eq. (\ref{eq:aux5}) is
\begin{equation*}
\begin{split}
& \int\limits_{Q(x_0)} \biggl(\nabla u, \nabla \frac{\dc u}{\dc x}\biggr) dx d\y =
\frac{1}{2} \int\limits_{Q(x_0)} \frac{\dc}{\dc x} (\nabla u, \nabla u) dx d\y \\
& = \frac{1}{2} \int \limits_{\Gamma(x_0)} (\nabla u, \nabla u) (\e_x, \n) dS + 
\frac{1}{2} \int \limits_{\Omega(x_0)} (\nabla u, \nabla u) d\y . \\
\end{split}
\end{equation*}

Taking into account the Dirichlet boundary condition $u|_{\Gamma(x_0)}
= 0$, one has
\begin{equation*}
\begin{split}
(\nabla u)|_{\Gamma(x_0)} & = \n \frac{\dc u}{\dc n}|_{\Gamma(x_0)} , \\
\frac{\dc u}{\dc x}|_{\Gamma(x_0)} & = (\e_x, \nabla u) = (\e_x, \n) \frac{\dc u}{\dc n}|_{\Gamma(x_0)} . \\
\end{split}
\end{equation*}
Combining these relations, one gets
\begin{equation*}
\begin{split}
& \int\limits_{Q(x_0)} \frac{\dc u}{\dc x} \Delta u ~dx d\y =
\frac{1}{2} \int\limits_{\Gamma(x_0)} \biggl(\frac{\dc u}{\dc n}\biggr)^2 (\e_x, \n) dS \\
& - \frac{1}{2} \int\limits_{\Omega(x_0)}
\biggl(\frac{\dc u}{\dc x}\biggr)^2 d\y + \frac{1}{2} \int\limits_{\Omega(x_0)} (\nabla_{\pe} u, \nabla_{\pe} u) d\y ,  \\
\end{split}
\end{equation*}
from which and Eqs. (\ref{eq:aux4}, \ref{eq:aux6}) the Rellich's
identity (\ref{eq:Rellich}) follows.

\end{appendix}

\end{document}